\documentclass[smallextended]{svjour3}
\usepackage[utf8]{inputenc}
\usepackage{amsmath, amsfonts, amssymb}
\usepackage{graphicx}
\usepackage{caption}
\usepackage{subcaption}
\usepackage[affil-it]{authblk}
\usepackage[english]{babel}
\usepackage[round,authoryear,sort&compress]{natbib}
\usepackage[bookmarks=true,breaklinks=true,bookmarksnumbered=true,colorlinks,citecolor=blue,linkcolor=blue,naturalnames,nesting,unicode]{hyperref}
\usepackage[linesnumbered,ruled,vlined]{algorithm2e} 
\usepackage[top=3cm,left=3cm,bottom=3cm,right=3cm]{geometry}
\usepackage{tabularx}
\usepackage{rotating}

\title{Maximising Throughput In A Complex Coal Export System}

\author{Mateus Rocha de Paula \and Natashia Boland \and Andreas Ernst \and Alexandre Mendes \and Martin Savelsbergh}

\institute{
	Mateus Rocha de Paula \at
	Hunter Valley Coal Chain Coordinator\\
	45 Lambton Road, Broadmeadow NSW 2292, Australia \\
	\email{Mateus.Rocha@hvccc.com.au}
	\and Natashia Boland \at
	Georgia Institute of Technology\\
	North Ave NW, Groseclose 408, Atlanta, GA 30332, United States\\
	\email{Natashia.Boland@isye.gatech.edu}
	\and Andreas T. Ernst \at
	Monash University, School of Mathematical Sciences \\
	Clayton Campus, Melbourne, Australia \\
	\email{Andreas.Ernst@monash.edu}
	\and Alexandre Mendes \at
	The University of Newcastle, School of Electrical Engineering and Computer Science \\
	University Dr., Callaghan, NSW 2308, Australia \\
	\email{Alexandre.Mendes@newcastle.edu.au}
	\and Martin Savelsbergh \at
	Georgia Institute of Technology\\
	North Ave NW, Groseclose 314, Atlanta, GA 30332, United States\\
	\email{Martin.Savelsbergh@isye.gatech.edu}
}

\date{Received: date / Accepted: date}

\begin{document}
\maketitle
\begin{abstract}
The Port of Newcastle features three coal export terminals, operating primarily in cargo assembly mode, that share a rail network on their inbound side, and a channel on their outbound side. Maximising throughput at a single coal terminal, taking into account its layout, its equipment, and its operating policies, is already challenging, but maximising throughput of the Hunter Valley coal export system as a whole requires that terminals and inbound and outbound shared resources be considered simultaneously. Existing approaches to do so either lack realism or are too computationally demanding to be useful as an everyday planning tool. We present a parallel genetic algorithm to optimise the integrated system. The algorithm models activities in continuous time, can handle practical planning horizons efficiently, and generates solutions that match or improve solutions obtained with the state-of-the-art solvers, whilst vastly outperforming them both in memory usage and running time.
\end{abstract}

\section{Introduction}

The Port of Newcastle features three coal export terminals coordinated by the Hunter Valley Coal Chain Coordinator (HVCCC): the Kooragang Coal Terminal (KCT) and the Carrington Coal Terminal (CCT), operated by the Port Waratah Coal Services (PWCS), and the Newcastle Coal Infrastructure Group (NCIG) Coal Terminal (NCT). Together, these three terminals are responsible for the largest coal exporting operation worldwide by tonnage, with a throughput of over 160 million tonnes in 2014. Thirty five coal mines, as far as 380 kilometers from the harbour, are connected to the terminals by a rail transportation system employing more than 40 coal trains and feeding over 1,600 coal vessels per year.

\begin{figure}[t!]
\centering
\includegraphics[scale=0.55]{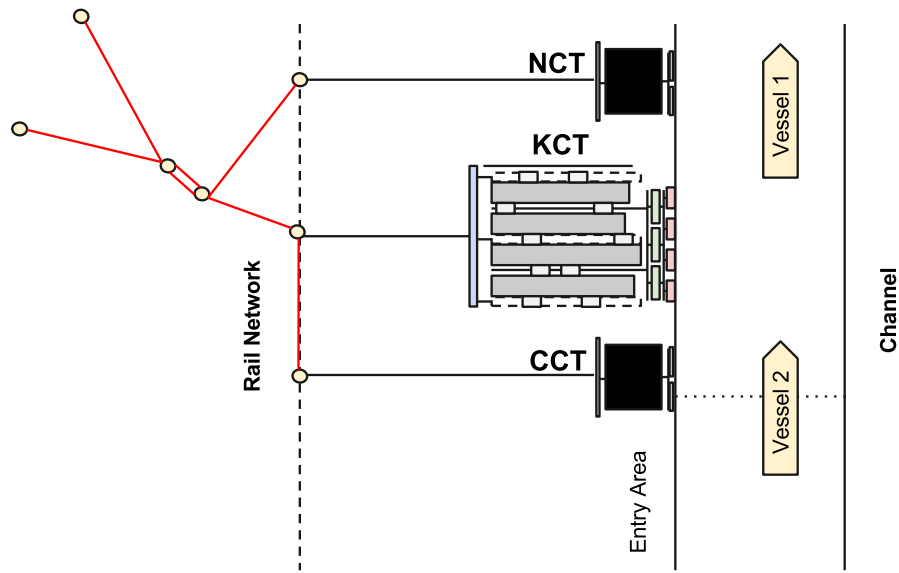}
\caption{The Hunter Valley Coal Chain (HVCC) has three coal terminals: Kooragang (KCT), Carrington (CCT) and NCIG (NCT). A shared rail network is used to transport coal from mines to terminals and a channel is used by vessels to transport coal to final customers. (Operations in KCT are modelled in more detail than in CCT and NCT.)}
\label{fig:system}
\end{figure}

The terminals share, on their inbound side, a rail network that connects them to the mines' load points, and, on their outbound side, a channel that connects them to the Pacific Ocean.  Most of the mines in the Hunter Valley are open pits, where coal is mined and stored either at a railway siding located at the mine or at a coal loading facility that can be shared. It is then transported to one of the terminals at the Port of Newcastle, almost exclusively by rail, dumped at their dump stations and then stacked on a pad to form stockpiles. Coal extracted from various mines, with different characteristics, is mixed into blended stockpiles to meet particular customers' specifications. Once a ship berths at a terminal, the appropriate stockpiles are reclaimed and loaded onto it. When fully loaded, the vessel may depart to its destination. \autoref{fig:system} illustrates the system under consideration.

The channel of the Port of Newcastle is quite narrow and shallow. For that reason, channel traffic must follow strict rules and procedures to avoid vessel clashes and damage to the ships' hull due to contact with the bottom of the channel, specially for bigger vessels. That, along with limited availability of outbound resources (i.e., number of berths and ship loaders) significantly limits outbound movements at the terminals.

KCT and CCT operate as Cargo Assembly (CA) coal loading terminals. That means that they work in a ``pull-based'' manner, where the coal blends are assembled and stockpiled based on the demands of the arriving ships. Ideally, for this operation mode, the assembly of the stockpiles for a vessel completes at the time the vessel arrives at a berth (i.e., just-in-time assembly) and the reclaiming of the stockpiles commences immediately. Unfortunately, this does not always happen due to the limited capacities of the resources in the system, such as stockyard space, availability of stackers and reclaimers, in/outbound capacities and channel availability. NCT can operate in a ``push-based'' manner, where coal is kept pre-blended for longer periods in dedicated stockpiles, owned by specific customers and kept at fixed locations on the NCT stock pads, and is reclaimed and loaded when demand appears. The majority of NCT customers, i.e., mining companies, however, operate in CA mode in their contracted space.

Regarding previous works focused on the HVCCC setting, \cite{Savelsbergh2014} consider Stockpile Location and Reclaimer Scheduling (SLARS) operations at KCT, accounting for pad assignment and placement, and reclaimer assignment with clash avoidance. \cite{Boland2011, Boland2012stockyard} also attempt to solve SLARS combining construction and mixed integer programming (MIP) based heuristics. These works, however, do not account for the shared resources consumed by coal passing through CCT and NCT, i.e., railing or channel usage.
\cite{Thomas2013}, on the other hand, attempt to control the integrated
system, with a single terminal and shared inbound and outbound resources,
using a distributed algorithm based on Lagrangian relaxation. Their work
models the operations at the terminal in a greatly simplified way and does not
consider channel traffic while including a more sophisticated model of the
rail operations. \cite{Belov2014} also tackles the integrated system, including in-terminal operations, coal arrival scheduling and channel traffic rules, using constraint programming. Since \cite{Thomas2013} and \cite{Belov2014} use time-indexed models, their approaches require a sufficiently high granularity to be of practical interest. Since the time slots are typically smaller than one hour, and the planning horizons under consideration are typically of the order of weeks, such methods tend to be very computationally demanding. Also, because they rely on external (commercial) optimisation solvers, their performance is dependent on the chosen (external) solver, and their use expensive.

This work describes a method that simultaneously schedules coal arrivals at the dump stations, determines build and load periods, and schedules arrival and departure times of the vessels obeying simple channel traffic rules. For KCT, which is the key terminal of the system and responsible for handling two thirds of the volume exported, our method also decides stockpile locations and schedules stockpile reclaiming (avoiding reclaimer clashes). The objective is to maximise the system's throughput without causing unacceptable vessel delays, which is one of the main challenges faced by the HVCCC. Other interesting and related problems are described in \cite{Boland2012optimising}.

The centre piece of our method is an enumerative algorithm to solve the SLARS subproblems for each terminal, which is based on the work of \cite{Savelsbergh2014}. It extends their algorithm by introducing rail network and channel considerations and accounting for CCT and NCT as shared resources consumers. A Parallel Genetic Algorithm is then used to introduce solution diversity and improve solution quality. Time is considered in a continuous fashion, and practical planning horizon sizes are handled efficiently. In addition, our approach does not rely on any external solvers, making it cost effective. The proposed algorithms generate very competitive solutions that match or improve solutions produced by state-of-the-art solvers, vastly outperforming them in terms of computational resources (i.e., both in memory usage and running time).

The remainder of this paper is organised as follows: \autoref{sec:problem} details the HVCCC system, our assumptions, and the problem under consideration. \autoref{sec:proposedmethods} elaborates on the methods proposed in this work: \autoref{sec:slars} details an extended version of the greedy algorithm proposed by \cite{Savelsbergh2014} and \autoref{sec:ga} describes a Genetic Algorithm that exploits this method to obtain better solutions. \autoref{sec:results} describes our experimental design to test the performance of our methods and \autoref{sec:conclusions} summarises our conclusions.

\section{The HVCC}
\label{sec:problem}

The three terminals in the Port of Newcastle share a rail network that connects the mines' load points to the terminals, and a channel that allows the ships to access the terminals. In this work, we optimise system-wide operations, which includes the terminals' inbound and outbound coal flows, taking into account critical restrictions at a sufficient level of detail to ensure that solutions of practical interest are obtained.  These critical restrictions are discussed below.


\subsection{The Rail Network}

\begin{figure}[t!]
\centering
\includegraphics[scale=0.35]{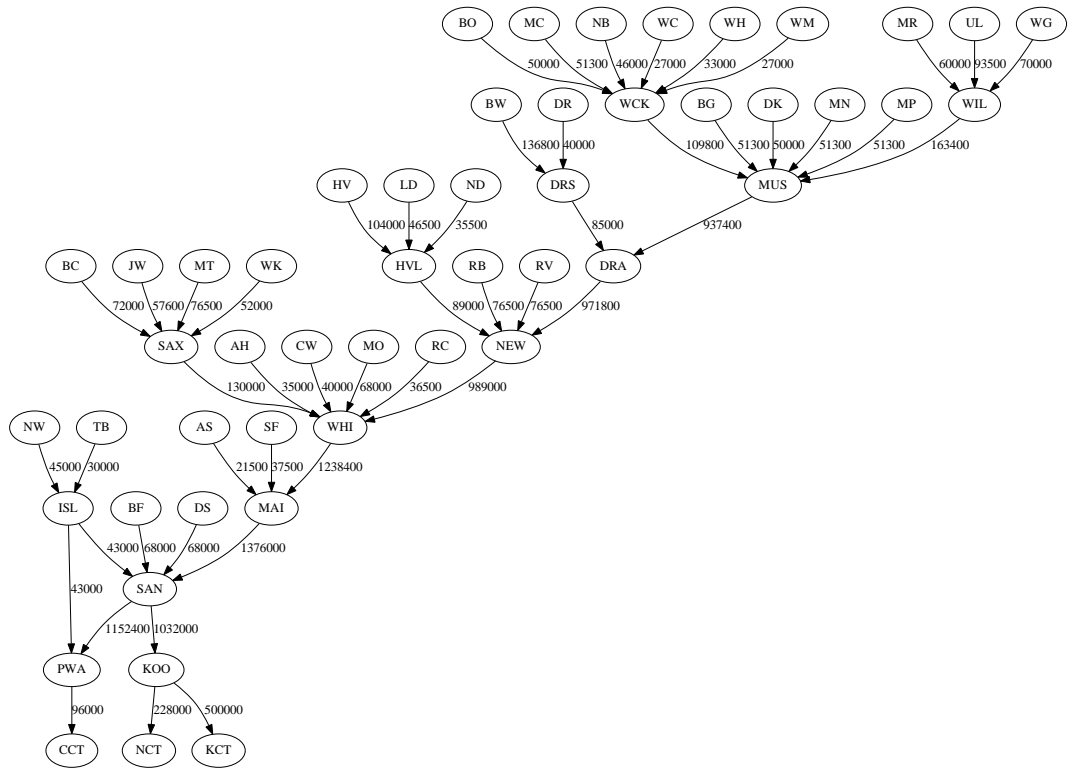}
\caption{The rail network is modelled as a directed graph. Coal flows from the load points to the terminals following a unique path.  When there are multiple paths, e.g., NW to CCT, the shortest path is always used. Travel times are ignored and coal is immediately available at the terminals as long as sufficient rail capacity along a path is available. Rail capacities are specified in terms of tonnes per day, and shown as arc labels.}
\label{fig:rail_network}
\end{figure}

The rail network is the largest part of the supply chain infrastructure,
connecting the train load points at open cut mines in the Hunter Valley to the
terminals at the Port of Newcastle.
For use in our methods, the rail network is modelled as a directed graph, with tonnes of coal flowing from the load points to the terminals along a unique path. In the few cases where there are multiple paths, the shortest path (in terms of number of arcs used) is always preferred. 
Each arc of the graph represents a relevant rail segment, and has a capacity given in tonnes per day. In practice, coal for a single stockpile is transported from a load point to a terminal using trains of a specific size, and rail capacity is given in both number of trains  and tonnes per day. However, for simplicity, we choose to schedule tonnes of coal to be delivered daily, without accounting for the number of trains. Scheduling of actual trains is delegated to the above-rail operators and also includes scheduling of crews, fueling, and maintenance. \autoref{fig:rail_network} depicts the relevant rail segments of the Hunter Valley.
Travel times are also ignored: coal is assumed to be immediately available at the terminals at the requested times as long as enough rail capacity for the path under consideration is available.

\subsection{The Channel}

The channel in the Port of Newcastle is quite narrow and shallow, it ranges
between about 330 and 670 meters in width and is only about 15m deep in some areas. For that reason, channel traffic must follow strict rules.

\begin{figure}
\centering
\includegraphics[scale=0.45]{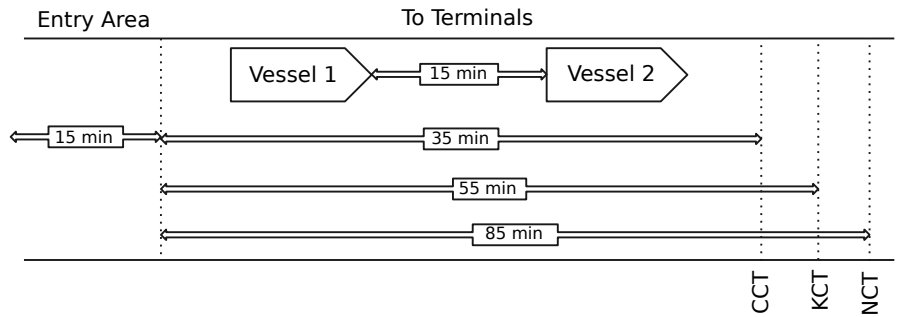}
\caption{The channel is modelled as a straight line and vessels are assumed to travel through the channel at the same constant speed. The channel has an entry area that takes 15 minutes to cross and three terminals, CCT, KCT and NCT, that can be reached in 35, 55 and 85 minutes from the end of the entry area, respectively. Traffic on the channel can only occur in one direction at a time and vessels must be at least 15 minutes apart. \label{fig:channel}}
\end{figure}

First of all, traffic can only happen on the channel in one direction at a time. Vessels are assumed to travel all at the same constant speed, and must be at least 15 minutes apart from each other. Departures and arrivals at a terminal are allowed to happen simultaneously.

Since the channel is relatively narrow, maneuvering on it requires the assistance of tug boats. Therefore, to account for staff limitations, we assume that at most four vessels can be travelling on the channel at any given time.

To avoid damaging their hull due to contact with the rocky bottom of the channel, large vessels (i.e.: with at least 100000t, also referred to as cape sized) can only depart during a tidal window [HT-90min, HT+30min), where HT is the time of the high tide. In this work, we consider real high tide times obtained using third party software\footnote{http://www.arachnoid.com/JTides/}. Vessels are considered to be capes if their requested load is at least 100kt.

Finally, we assume that vessels take 15 minutes to travel through the
channel's entry area, and 35, 55 and 85 minutes to travel from its end to CCT,
KCT and NCT, respectively. The reverse trips also need to be allowed for and
take the same amount of time. \autoref{fig:channel} depicts the relevant channel information.

\subsection{The Terminals}

In order to obtain realistic solutions that can be used in practice by the HVCCC, certain key aspects, rules and restrictions associated with the system must be observed.

First of all, in any of the terminals, the amount of coal being dumped at the stations in a single day cannot exceed the Daily Inbound Throughput (DIT). Similarly, the amount of coal loaded into vessels cannot exceed the Daily Outbound Throughput (DOT).

Next, split cargoes are typically seen as an undesirable occurrence, and therefore forbidden for the purposes of our methods. In other words, one cargo equals one stockpile. Also, vessels must be loaded in such a way that they maintain their physical balance in the water. Therefore, their stockpiles must be reclaimed in a pre-specified order.

In this work, we distinguish KCT from the other two terminals because it is responsible for the largest volume of coal exported, with bigger inbound and outbound capacities and availability of equipment; making it a key terminal for the system. That also means that optimising operations in the terminal to meet the demands is itself a challenging problem. NCT is responsible for a significant volume of coal exported as well, but HVCCC has only limited visibility and control over their operations, and, as a result, creating a detailed plan is impossible and not useful in practice. CCT is also modelled in coarse detail because only a small volume of coal that passes through it.  However, CCT needs to be considered as it does share the rail network and channel with the other two terminals, which are resources with limited capacity.

While for KCT we consider operations in the terminal in detail, for CCT and NCT we only aim to schedule coal deliveries, vessel arrivals and departures, and stockpile build and reclaim periods. How the stockpiles are positioned on the pads, and which equipment is used to build and reclaim the stockpiles is left for the planner to decide.

Since we propose a day-of-execution plan for cargoes in every terminal, to avoid exceedingly long build times, stockpiles can start their stack periods at most ten days prior to their vessels' Estimated Arrival Time (ETA). Also, because train loads often cannot be delivered at the ideal time due to limited rail capacity, stacking can be preempted (i.e., days with no stacking are allowed after the stacking period has started). Exceedingly large build times are also undesirable, as the stockpiles would occupy scarce pad space that could be used for other stockpiles. Therefore, build periods are restricted to a maximum of seven days. Finally, since train travel times are not considered in our model (i.e., coal is considered to be immediately available at the terminal upon request, subject only to rail capacity and DIT), unrealistically short stack periods are avoided by enforcing a minimum build period of three days. Maximum build times can also depend on the mines that provide coal, e.g., because some of the mines are relatively close to the terminal, a 5 day limit is sufficient and more appropriate.

Reclaiming a stockpile can only start once it is fully built to avoid that reclaiming has to be interrupted because of lack of coal (building the stockpile had not yet finished), forcing a vessel to remain berthed without being loaded. Such a situation is highly undesirable since berths are a very scarce resource.

Since manoeuvring in a narrow channel can be difficult and requires resources that could be used for other purposes, once a ship berths at a terminal, it cannot change berths. However, it is easy to see that, as long as the number of vessels berthed never exceeds the number of berths in the terminal, a first-come-first-served policy ensures that this requirement is satisfied.

\subsubsection{The Kooragang Coal Terminal}

\begin{figure}[t!]
\centering
\includegraphics[scale=0.35]{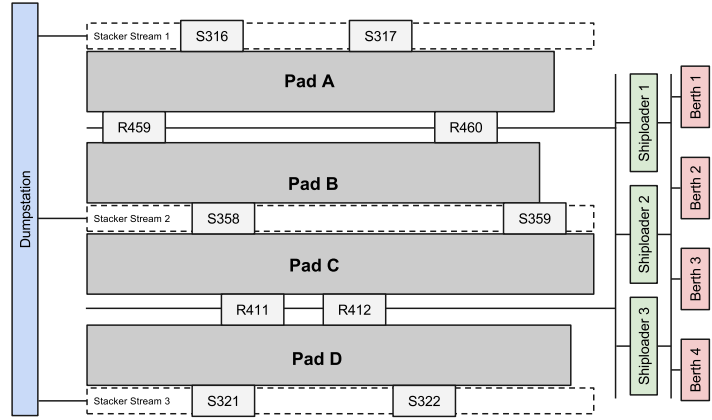}
\caption{The Kooragang Coal Terminal is modelled in more detail than the other two terminals. There are four pads, A, B, C and D, with known lengths 2142m, 1905m, 2174m and 2156m, respectively, four reclaimers, R459 and R460 operating on pads A and B and R411 and R412 operating on pads C and D, and six stackers, S316 and S317 operating on stacker stream 1, S358 and S359 operating on stacker stream 2, and S321 and S322 operating on stacker stream 3. Stacker stream 1 serves Pad A, stacker stream 2 serves pads B and C and stacker stream 3 serves pad D. Stacker streams 1 and 3 are single railed and have a capacity of 144kt per day, while stacker stream 2 is double railed and has twice this capacity.  KCT can accommodate an inbound flow of 500kt per day, has three ship loaders and four berths which, combined, can accommodate an outbound flow of 390kt per day.}
\label{fig:kct}
\end{figure}

The stockyard at KCT has four pads, A, B, C, and D, on which cargoes are assembled. Upon arrival at the terminal, a train dumps its contents into one of three stations. The coal is then transported on a conveyor to one of the pads where it is added to a stockpile by a stacker. Since all dump stations can send coal to any stacker stream, in this work we model them as a single dump station with the combined capacity. There are six stackers, two that serve pad A (S316 and S317), two that serve pad B and pad C (S358 and S359), and two that serve pad D (S321 and S322). A single stockpile is built from several train loads over three to seven days, as mentioned in the previous section. After a stockpile is completely built, it may dwell on its pad for some time until its destination vessel has arrived. Stockpiles are reclaimed using a bucket-wheel reclaimer and the coal is transferred to one of the four berths on a conveyor. The coal is then loaded onto the vessel by a ship loader. There are four reclaimers, two that serve pad A and pad B (R459 and R460) and two that serve pad C and pad D (R411 and R412). \autoref{fig:kct} illustrates KCT's layout, as modelled in this work.

Pads A, B, C and D are 2142m, 1905m, 2174m and 2156m long, respectively, and we assume that every stockpile occupies the full width of the pad. \autoref{eq:tonnes2metres}, obtained by linear regression from actual data, is used to approximate the length of a stockpile, under this assumption, given its tonnage $t$. Naturally, two stockpiles cannot occupy the same space on a pad at the same time (i.e., coal cannot be shared between stockpiles). Also, once the assembly of a stockpile has started, it is rare that the location of the stockpile in the stockyard is changed. Relocating is time-consuming and requires resources that could be used to assemble or reclaim other stockpiles. For this reason, this is forbidden for our purposes. We also assume that the terminal has a DIT of 500kt/day.

\begin{equation}
\label{eq:tonnes2metres}
5.0\Bigl\lfloor\frac{0.0017t+39.714}{5.0}+0.5\Bigr\rfloor
\end{equation}

We assume that stackers on KCT move at a speed of 1800m/h and operate at a rate of 139.2kt/day, which makes stacker travel and operation times (of the order of minutes) insignificant compared to train cycle times (of the order of days). For this reason, we do not model stacker operations. It is assumed that it will always be possible and straightforward to build the stockpiles within the stipulated stacking period, provided that the terminal's capacities are respected, and this task is left for the planner. The middle stacker stream has two conveyor belts and has a Daily Stacker Stream Capacity (DSSC) of 288kt/day. That is twice the capacity of the outer stacker streams, which only have one conveyor belt.

We assume that reclaimers at KCT also move at a speed of 1800m/h and operate at a rate of 139.2kt/day. Reclaimers that serve the same pads cannot pass each other, as they travel on rails on the side of a pad. Reclaimers can only load coal from one stockpile at a time, and can only be assigned to stockpiles on pads that they serve. Also, since the terminal has only three ship loaders, only three reclaimers can operate at the same time. Reclaim jobs are not preemptive and, since there is a limiting number of berths, it is desirable that a vessel is never idle whilst berthed and occupying precious space. Therefore, once loading of a vessel has started, it cannot be stopped (in between stockpiles) for more than 5 hours. The combined total reclaimer and ship loader capacities enforce a DOT of 390kt/day.

\subsubsection{The NCIG and Carrington Coal Terminals}

\begin{figure}
\centering
\includegraphics[scale=0.35]{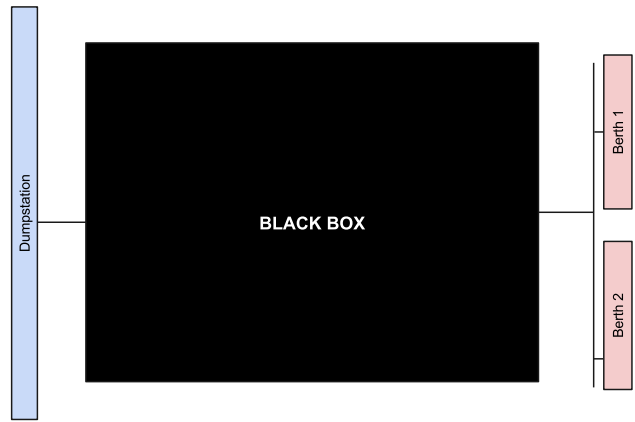}
\caption{The Carrington and NCIG Coal Terminals are modelled in less detail than KCT. Operational aspects related to the terminal, other than the interfaces with the shared components, i.e., the rail network and the channel, are not modelled. CCT can accommodate an inbound flow of 96Mt per day while NCT can accommodate an inbound flow of 228kt per day. The outbound interface is simplified to the number of berths and a maximum daily outbound flow (94 and 214kt for CCT and NCT, respectively).}
\label{fig:cct_nct}
\end{figure}

In this work, we do not model stockyard operations in CCT and NCT. Even though NCT is responsible for a significant volume of coal exported, HVCCC has only limited visibility and control over their operations as NCIG customers, who have contracted dedicated stockpile space at the terminal typically do not disclose their operational data, which makes detailed modeling impossible and impractical. CCT is also modelled in coarse detail because only a small volume of coal that passes through it. However, it is important to consider it in the integrated system because it shares the rail network and channel with the other two terminals, which are limited capacity resources. \autoref{fig:cct_nct} illustrates CCT's and NCT's layout, as modelled in this work.

Even though these terminals support dedicated stockpiles, it is known that most of the real estate owners at the terminals operate their stockpiles in CA mode. Therefore, we approximate their operations under this assumption, and do not model dedicated stockpiles at all.

It is assumed that CCT has a DIT of 96kt/day and a DOT of 94kt/day. NCT is assumed to have a DIT of 228kt/day and a DOT of 214kt/day. CCT features two berths only while NCT has three. We also assume maximum reclaim rates of 2200t/h and 5800t/h for CCT and NCT, respectively.

\subsection{Shipping Stems}

In this work, shipping stems are used to characterise the input data. A shipping stem is a list of vessels with information on their ETA, the terminal they are headed to, and their cargo details. Cargoes are specified in terms of their coal components, i.e., each cargo specification consists of a list of components with for each component a tonnage and the load point of origin.

\section{Algorithms to Optimise the Integrated System}
\label{sec:proposedmethods}

This work describes a method that simultaneously schedules coal arrivals at the dump stations, determines stockpile build and load periods, and schedules arrival and departure times of the vessels. In the case of KCT, which is a key terminal of the system, the method also determines a stockpile placement (pad plus location) and assign a reclaimer (accounting for the fact that two reclaimers operating on the rail track cannot pass each other to reach their designated stockpiles). \autoref{tab:notation} summarises the notation used throughout this paper.

\begin{table}
\centering
\textbf{Data}

\begin{tabularx}{\linewidth}{lX}
	\hline
	$V = \{v_1 ... v_{|V|}\}$ & The set of vessels.\\
	$ETA_v, \forall v \in V$ & The Estimated Time of Arrival (ETA) of a vessel.\\
	$\underline{d_v}, \forall v \in V$ & The earliest possible departure time of a vessel.\\
	$S_v = \{s_{v,1} ... s_{v,|S_v|} \} \forall v \in V$ & The set of stockpiles of a vessel.\\
	$C_s = \{c_{s,1} ... c_{s,|C_s|}\}$ & The set of components of a stockpile.\\
	$S = \bigcup_{v \in V} S_v$ & The set of stockpiles.\\
	$w_v = \sum_{s \in S_v} w_s, \forall v \in V$ & The tonnage (weight) of a vessel (the total tonnage of its stockpiles).\\
	$w_s = \sum_{c \in C_s} w_c, \forall s \in S$ & The tonnage (weight) of a stockpile (the total tonnage of its components). \\
	$w_c \forall c \in C_s$ & The tonnage (weight) of a component.\\
	$R = {r_1 ... r_{|R|}}$ & The set of rail segments, i.e., the arcs in the graph shown in \autoref{fig:rail_network}. For simplicity, throughout this paper, we may also refer to a rail segment by its descriptor rather than its index (i.e.: $r_i = MUS$).\\
	$\overline{w_r} \forall r \in R$ & The capacity of a rail segment.\\
	$H = {h_1 ... h_{|H|}}$ & The set of high tides.\\
	$W_h = [h-1.5, h+0.5) \forall h \in H$ & The set of tidal windows. Note that, throughout this work, time is given in hours and we may also refer to the limits of the tidal window as $[\underline{W_h}, \overline{W_h})$.\\
	$W_t = W_h : \{\overline{W_h} < t\}_{1st}$ & The tidal window that
        contains a time $t$ or the first tidal window after time $t$, if it is
        not contained in any tidal window.\\
        $L$ & The highest reclaime rate of a terminal (5800 t/hr except for CCT
        which as a limit of 2200 t/hr)\\
\end{tabularx}

\vspace{0.5cm}
\textbf{Decision variables}
\flushright
\emph{All terminals}
\begin{tabularx}{\linewidth}{lX}
	\hline
	$D$ & The average vessel delay as defined by \autoref{eq:delay}.\\
	$a_v \forall v \in V$ & Arrival time of vessel $v$.\\
	$d_v \forall v \in V$ & Departure time of vessel $v$.\\
	$\tau_s = \{(w, t)\} = \bigcup_{c \in C_s} \tau_c \forall s \in S$ & The set of coal arrivals for a stockpile (the set of coal arrivals for each of its components).\\
	$\tau_c = \{(w, t)\} \forall c \in \bigcup_{s \in S} C_s $ & The set of coal arrivals for a component. A coal arrival specifies the tonnes $w$ delivered. Note that, since inbound capacities are given on a daily basis and we model time on a hourly basis, $t$ is always a multiple of 24 here. Note also that $\sum_{\tau_c=\{w,t\}}w=w_c$.\\
	$b_s = \{\underline{b_s}, \overline{b_s}\} \forall s \in S$ & The stacking (building) period of a stockpile. Note that, throughout this paper, $\underline{b_s} = min_{\tau_s=\{w,t\}}t$ and $\overline{b_s} = max_{\tau_s=\{w,t\}}t$.\\
	$l_s = \{\underline{l_s}, \overline{l_s}\} \forall s \in S$ & The reclaiming (loading) period of a stockpile.\\
	$l_v  \forall v \in V$ & The reclaiming (loading) period of a vessel. Since the stockpiles have to be reclaimed in order, $\underline{l_v} = \underline{l_{s_{v,1}}}$ and $\overline{l_v} = \overline{l_{s_{v,|S_v|}}}$.\\
\end{tabularx}

\vspace{0.25cm}
\emph{KCT only}
\begin{tabularx}{\linewidth}{lX}
	\hline
	$P_s \in \{Pad A .. Pad D\} \forall s \in S$ & The pad on which a stockpile will be built.\\
	$p_s \forall s \in S$ & The position on the pad (length, in metres, from $l=0$) where the \emph{middle} of the stockpile will be when built.\\
	$R_s \in \{R411, R412, R459, R460\} \forall s \in S$ & The reclaimer that will load the stockpile on its vessel.\\
	$R_P$ & The reclaimers that service pad $P$. In this work, $R_{PadA} = R_{PadB} = \{R459, R460\}$ and $R_{PadC} = R_{PadD} = \{R411, R412\}$.\\
	\hline
\end{tabularx}
\caption{Summary of notation}
\label{tab:notation}
\end{table}

Our objective is to maximise the system's throughput without causing unacceptable vessel delays. We define the earliest departure time $\underline{d_v}$ for a vessel $v$ as the ideal departure time: as if the vessel could berth at the terminal exactly at its ETA, had all its stockpile pre-built, had a reclaimer ready to load it without interruptions, and could depart immediately after its fully loaded (or at the very beginning of the next high tide, in the case of capes). Let $\underline{H_t}$ be the beginning of the first tide that happens during or after time $t$, $L$ be the highest reclaim rate of a terminal (i.e.: 5800t/h for KCT and NCT, and 2200t/h for CCT) and $w_v$ the total tonnage of the vessel. \autoref{eq:earliestDepartureTime} formulates this concept.

\begin{equation}
\underline{d_v} = 
\left \{
	\begin{tabular}{ll}
	$max\{ETA_v + Lw_v, \underline{H_{ETA_v+Lw_v}} \}$ &, if $v$ is a capesize type vessel \\
	$ETA_v + Lw_v$ &, otherwise
	\end{tabular}
\right. 
\label{eq:earliestDepartureTime}
\end{equation}

The system's average vessel delay, formulated by \autoref{eq:delay}, is used as a proxy for maximising the throughput, where $d_v$ denotes the vessel's actual departure time.

\begin{equation}
\frac{\sum_{v \in V}{d_v - \underline{d_v}}}{|V|}
\label{eq:delay}
\end{equation}

\subsection{A Greedy Algorithm for Stockyard Management}
\label{sec:slars}

The centre piece of our method is an enumeration algorithm to solve the
stockyard management problem at a terminal, e.g., the placement, the stacking,
and the reclaiming of stockpiles, captured in the SLARS sub-problems for each
terminal. The procedure sequentially schedules vessels in a greedy fashion:
once a vessel and its stockpiles are scheduled, this decision is only
revisited if a feasible solution based on it can not be found.
\autoref{algo:slars}() details this method. The important question of how to find a good
input sequence to the SLARS sub-problems is discussed in Section~\ref{sec:ga} below.

In this work, we only work with \emph{feasible} solutions, in the sense that every vessel schedule and stockpile placement satisfies every constraint described in \autoref{sec:problem}. We refer to an \emph{incomplete} solution as one in which not all vessels have been scheduled or stockpiles have been placed (yet).

\emph{Scheduling a vessel $v$} refers to setting the arrival and departure
times $a_v$ and $d_v$, respectively. Since a vessel can only depart once it is
fully loaded, that can only be done after all its stockpiles are placed.
\emph{Placing a stockpile $s$}, refers to setting the build and loading
periods $b_s$ and $l_s$, respectively. In order to build a stockpile, we first
schedule the delivery of coal for all its components. The latter is referred
to as \emph{railing}, and refers to determining the coal arrivals $\tau_c$ for
every component $c \in C_s$. We assume that each component comes from a single
unique mine, via an unique path on the graph shown in
\autoref{fig:rail_network}. Since reclaiming of a stockpile can only start
after it is fully built, the reclaiming period can start only after the
railing has been completed. The details for scheduling a vessel and placing
the stockpile are given as pseudo-code. Note that \ref{algo:getPadGaps}
considers all stockpiles from all ships that are already placed. However the
backtracking never needs to go back further than the current ship, as it is
always possible to place all stockpiles at the end of the time axis.

\begin{procedure}
\scriptsize
$Solution \leftarrow \emptyset$\;
\ForEach{$v$ in $V\prime$}{
	\lIf{$v$ goes to $KCT$} {
		\ref{algo:kctscheduler}($v$)
	}
	\lElse{
		\ref{algo:stdscheduler}($v$)
	}
}
\Return{Solution}
\caption{slars(order $I$): given a pre-specified evaluation order, this procedure sequentially and greedily schedules vessels, setting the vessel's decision variables considering only the decisions made up to that point and these choices are never revisited.}
\label{algo:slars}
\end{procedure}

\begin{procedure}
\scriptsize
\ForEach(\texttt{//In the order they are defined}){Stockpile $s \in S_v$}{
	$(\tau_s, b_s) \leftarrow \ref{algo:railing}(s)$\;
}
\tcp{All stockpiles will be loaded non-preemptively, starting when they can, after the last stockpile is fully built}
$(l_v, a_v, d_v) \leftarrow$ \ref{algo:loadingPeriod}($v$)\;
\caption{scheduleVessel(Vessel v): a simplified vessel scheduler. Schedules coal arrivals for the stockpiles as early as possible, ignoring operational aspects related to the terminal, i.e., machine assignment and positioning; schedules reclaim jobs for the vessel, assuming that stockpiles are fully built; and determines vessel arrival and departure times that allow such schedules.}
\label{algo:stdscheduler}
\end{procedure}

\begin{procedure}
\scriptsize
\ForEach{$s \in S_v$}{
	$placements_s \leftarrow \emptyset$\;
	\tcp{Enumerate (and remember) all possible placements for every stockpile of this vessel}
	\ForEach{Pad $P_s \in \{Pad A .. Pad D\}$}{
		\ForEach{$padGap \in$ \ref{algo:getPadGaps}($P_s, s$)}{
			$\tau_s \leftarrow \ref{algo:railing}(s)$ \tcp{$b_s$ is now automatically known}
			\ForEach{Reclaimer $R_s \in {R_p}$}{
			 	\ForEach{$reclGap \in $ \ref{algo:getReclGaps}($R_s, s$)}{
			 		\ForEach{$criticalHeight \in$ \ref{algo:getCriticalHeights}($padGap,reclGap$)}{
			 			$l_s \leftarrow \ref{algo:getEarliestReclaimTime}(\overline{b_s}, padGap, reclGap, criticalHeight)$\;
						$placements_s \leftarrow placements \cup \{ (\tau_s, b_s, l_s) \}$\;
			 		}
			 	}
			}
		}
	}
	\tcp{Note that there will always be at least one placement for the first stockpile of the vessel (at the very end of the line, in the worst case), so we will never have to undo further than the first stockpile of the vessel (i.e.: undo a stockpile of another vessel)}
	\eIf{$placement_s == \emptyset$} { 
		undo the placement of the previous stockpile, use its next best placement and repeat this iteration\;
	}{
		sort $placement_s$ using \ref{algo:comparePlacements}()\;
		$(\tau_s, b_s, l_s) \leftarrow $ first (best) placement of $placement_s$\;
	}
	\If{$s$ is the last to be loaded onto $v$} {
		$(l_v, a_v, d_v) \leftarrow$ \ref{algo:loadingPeriod}($v$)\;
		\If{either $a_v$ or $d_v$ doesn't exist} { 
			regret the placement of the previous stockpile, use its next best placement and repeat iteration \;
		}
	}
}
\caption{scheduleVesselKCT(Vessel v): schedules vessels considering operational aspects of the terminal in detail. Every possible stockpile positioning and machine assignment is evaluated and the best one is chosen.}
\label{algo:kctscheduler}
\end{procedure}

\begin{procedure}
\scriptsize
starting at the latest time when one of $v$'s stockpiles is fully built, finds the first interval $[a_v, d_v)$ in which:
\begin{enumerate}
	\item there is at least one berth available throughout the whole period
	\item a vessel can traverse the channel and reach the terminal at $a_v$
	\item the vessel can depart the terminal at $d_v$ and traverse the channel
	\item there are less than four vessels traversing the channel throughout the whole period
	\item there is a sub-interval $l_v$ long enough to reclaim all the stockpiles in $v$ (i.e.: has enough DOT to do so during $[a_v, d_v)$, assuming that arbitrary tonnages can be reclaimed at any time).
\end{enumerate}
\tcp{In our implementation this is a search performed by sequential inspection.}
\KwOut{$a_v, d_v, l_v$}
\caption{getLoadingPeriod(Vessel $v$)}
\label{algo:loadingPeriod}
\end{procedure}

\subsubsection{Railing}

In this work, capacities are given on a per-day basis. We consider three such capacities: (1) rail segment capacity, (2) terminal inbound capacity (DIT) and, for KCT, (3) stacking capacity (DSSC).

Railing is always performed in a greedy fashion, for stockpiles and components, in the order the stockpiles have to be loaded into the vessels. There is no particular reason for this ordering (the stockpiles may be built in any order), but this one was chosen for simplicity. We assume that arbitrary amounts of coal can be transferred on any day, subject to available rail capacity. Train travel times are not considered: as long as there is enough capacity, we assume that coal flows instantaneously from a load point to a terminal. Therefore, a coal arrival $\tau_c = (w,t)$ for a component $c$ is to be interpreted as $w$ tonnes of coal arriving at a terminal on day $t$.

For a given stockpile, as many tonnes of coal as possible are scheduled to arrive as soon as possible, starting ten days before the vessel's ETA. \autoref{algo:railing}() details this approach.

\begin{procedure}
\scriptsize
\For{$c \in C_s$} {
	$TonnesLeft \leftarrow w_c$\;
	\tcp{In the case of KCT, stacking commences either at the beginning of the pad gap or ETA, whichever comes first}
	\lIf{$s$ goes to KCT} {
		$start \leftarrow \max\{ETA_v, pgs\}$
	}
	\lElse{$start \leftarrow ETA_v$}
	\For(\texttt{//Let $v$ be the vessel to which $s$ is destined}){$t \in [start, \infty)$}{
		$\alpha \leftarrow $ smallest rail capacity left on the path from the component's load point to its designated terminal\;
		$\beta \leftarrow $ DIT left on day $t$ for the designated terminal\;
		\lIf{$s$ goes to KCT}{$\gamma \leftarrow$ DSSC left on day $t$ for the stacker stream under consideration}
		\lElse(//Just a big constant otherwise){$\gamma \leftarrow M$}
		\tcc{$\alpha$, $\beta$ and $\gamma$ all take uncommitted capacity (i.e.: that will be reserved by the previous components of $s$) under consideration as well}
		$w \leftarrow \min\{\alpha, \beta, \gamma, TonnesLeft\}$\;
		\If{$w>0$} {
			$\tau_c \leftarrow \tau_c \cup (w, t)$\;
			$TonnesLeft \leftarrow TonnesLeft - w$\;
		}
		\lIf{$TonnesLeft=0$}{\textbf{break}}
	}
}
\KwOut{$\tau_s, b_s$}
\caption{railing(Stockpile $s$, $\lbrack$ $pgs \rbrack$): schedule as many coal arrivals as possible and as early as possible. Let $v$ be the vessel in which $s$ will be loaded. In the case of KCT, the start of the pad gap ($pgs$) in which $s$ is placed is also required.}
\label{algo:railing}
\end{procedure}

\subsubsection{Channel Traffic}

In this work, we assume that vessels travel at the same constant speed. Under this assumption, channel traffic can be controlled by selecting appropriate arrival and departure times.

Assuming that two vessels are headed to the same terminal, the following must hold whenever we refer to the possibility of a vessel traversing the channel, i.e., items (2) and (3) of \ref{algo:loadingPeriod}():
\begin{itemize}
	\item Any two consecutive arrivals must happen at least 15 minutes apart;
	\item Any two consecutive departures must happen at least 15 minutes apart; and
	\item If preceded by a departure, any arrival must wait until the departing vessel has cleared the channel. Therefore, two such consecutive events must be $2(t + 15)$ minutes apart, where $t$ is the time required for a vessel to traverse the channel from the entry area to its destination terminal.
\end{itemize}

\begin{figure}
\centering
\includegraphics[scale=0.3]{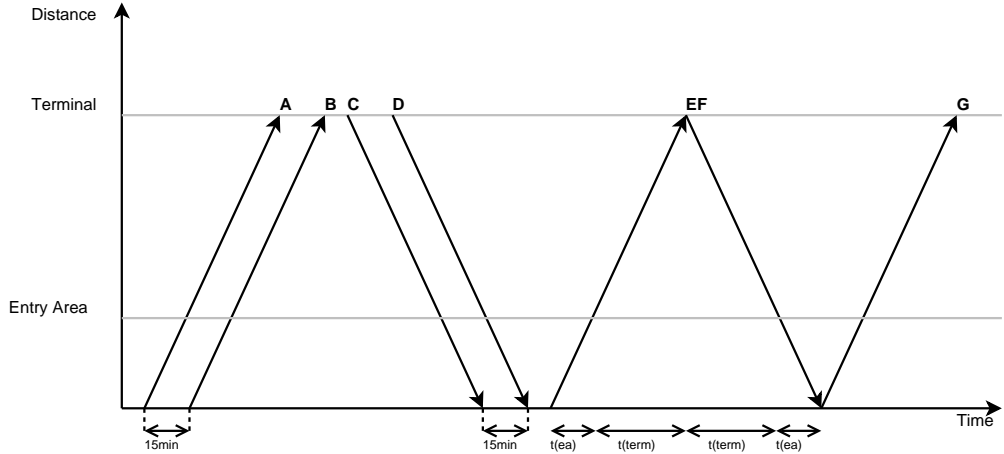}
\caption{Channel traffic: vessels can only travel the same section of the channel in one direction at any given time, and must be at least 15 minutes apart. In the figure above, vessels arrive at the terminal at times $A,B,E$ and $G$, and depart from it at times $C,D$ and $F$. }
\label{fig:channel_traffic}
\end{figure}

\begin{figure}[b]
\centering
\includegraphics[scale=0.4]{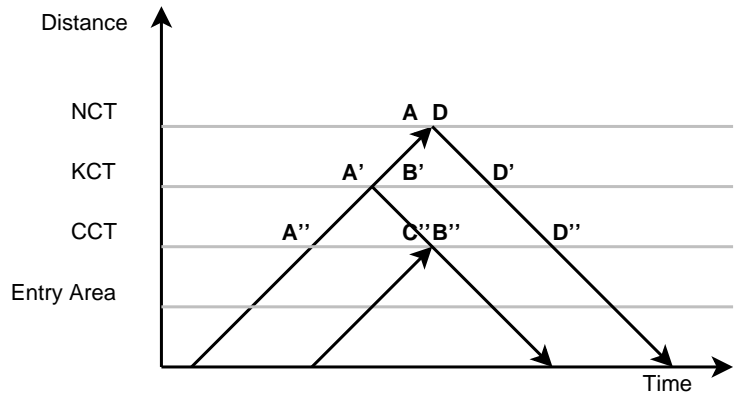}
\caption{Channel traffic with multiple terminals: we consider one time line for each terminal, and events (passing, arriving, and departing) are registered at every terminal encounter before reaching the destination terminal. In the above figure, a vessel arrives at NCT at time $A$, passing by KCT at $t=A'$ and CCT at $t=A''$. That allows another vessel to leave KCT at $t=B'=A'$ because the previous vessel won't travel on that section any more. This second vessel passes by CCT at $T=B''$, which allows another vessel to arrival at CCT $t=C''=B''$ because $C'' \geq A''+15min$.}
\label{fig:channel_traffic_mt}
\end{figure}

For simplicity, we assume that a departure may happen at the same time as a preceding arrival: as one vessel arrives at the berth, another leaves (if allowed by the rules above). \autoref{fig:channel_traffic} illustrates these rules.
Without loss of generality, these rules are easily extended to multiple terminals by keeping \emph{projected} events (arrival or departures) for each of the terminals. Under the assumption that the terminals are located along the channel, that the channel can be represented by a straight line, and the vessels travel at the same constant speed, an arrival is registered at a terminal every time a vessel berths at or passes by it on its way to its destination. Similarly, a departure is registered at a terminal every time a vessel departs from or passes by it on its way back. If the terminal where the event is being recorded is not the vessel's destination, i.e., it is a terminal encountered before the destination, the event time recorded is the time when the vessel passes the terminal. Naturally, each terminal's event list must satisfy the channel rules at all times. \autoref{fig:channel_traffic_mt} illustrates this idea.
Our methods implement these checks by sequential inspection. It is interesting
to note that CCT, the terminal that is most conveniently located on the
channel, is the oldest facility with low capacity and hence least used.

\subsubsection{Optimising Operations in KCT}

Unlike CCT and NCT, for vessels headed to KCT, we also determine a pad and position on the pad for stockpiles and schedule the time the stockpile will occupy the assigned space and when and by which reclaimer the stockpile will be reclaimed. Since there may be multiple ways of doing so, \autoref{algo:slars}() enumerates a finite set of possibilities and chooses the one that increases the total vessel delay the least. This is done by observing the geometrical aspects of the pads, in the same way as described by \cite{Savelsbergh2014}. This section describes the relevant procedures and we refer the reader to the original publication for more detailed information.

Under the assumption that the stockpiles occupy the entire width of a (rectangular) pad, the positioning of the stockpiles is uni-dimensional along the length of the pad. Consider a two dimensional plane with time along the horizontal axis and the position on a pad along the vertical axis. In this plane, a stockpile placement can be represented in time and space as a rectangle. In this representation, the coal ground period -- which includes stacking, reclaiming, and dwell periods -- is defined by horizontal boundaries of this rectangle on the section of the pad limited by its vertical boundaries.

\begin{figure}
\centering
\includegraphics[scale=0.6]{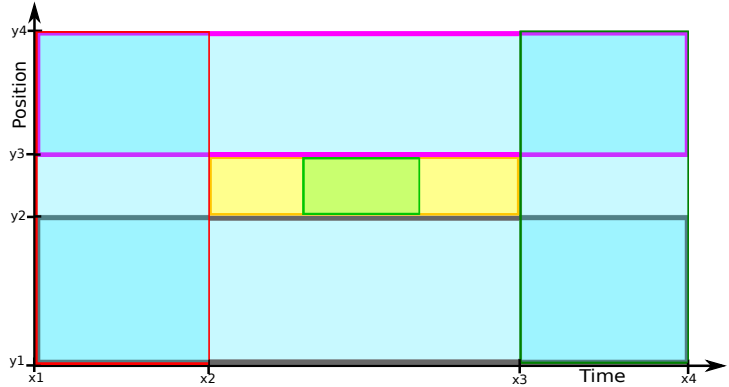}
\caption{Pad Gaps represent the available space and time on the stock pad for a given stockpile (to accommodate the stockpiles build, dwell, and load periods). In the figure above, suppose a single stockpile is already placed on the stock pad and is represented by the yellow rectangle (where the green area represents dwell time, the yellow area before represents build time, and the yellow area after represents load time). The area in blue represents the area in time and space where other stockpiles can be placed. Each of the four rectangles with maximal area ($(\langle x_1, y_1\rangle, \langle x_2, y_4\rangle), (\langle x_3, y_2\rangle, \langle x_4, y_4\rangle), (\langle x_1, y_1\rangle, \langle x_4, y_2\rangle)$ and $(\langle x_3, y_1\rangle, \langle x_4, y_4\rangle)$) that together represent this space is a pad gap. Note that the pad gaps may overlap each other (as indicated by the transparency around the edges of the pad gaps).}
\label{fig:padGaps}
\end{figure}

The available time and space to place other stockpiles is then the area that lies outside these rectangles. This area, in turn, can be divided in rectangles of maximal size, referred to as \emph{pad gaps}. A set of pad gaps is used as an algorithmically convenient structure to represent the available pad space. \autoref{fig:padGaps} illustrates this concept, which implemented by \autoref{algo:getPadGaps}().

\begin{procedure}
\scriptsize
\caption{getPadGaps(Pad $P$, Stockpile $s$): Returns the set of all (adjusted) pad gaps on pad $P$ that can accommodate building and loading of stockpile $s$ (assuming no dwell) based on the current state of the system/solution. See \autoref{fig:padGaps} for information about pad gaps.}
\label{algo:getPadGaps}
\tcp{In our implementation we start with a single pad gap that comprises the whole pad, and split it appropriately as stockpiles are placed.}
\KwOut{A set of pad gaps.}
\end{procedure}

\begin{figure}
\centering
\includegraphics[scale=0.6]{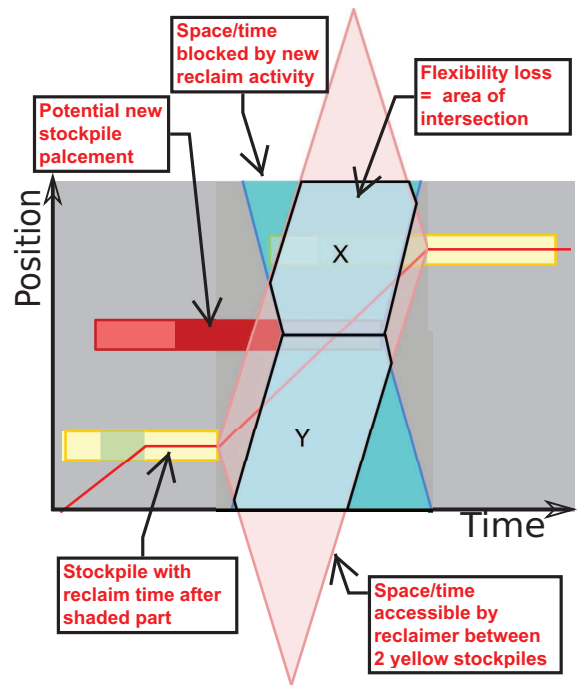}
\caption{Reclaimer gaps and flexibility loss. In the above figure, suppose the two stockpiles depicted in yellow (dwell period highlighted in green, with build and load periods immediately before and after dwell, respectively) are placed on a pad. Furthermore, assume that a single reclaimer loads both these stockpiles and moves at constant speed.  The movement of the reclaimer is illustrated by the red line.  The area highlighted in pink is the reclaimer gap: the area between two existing reclaim jobs of a reclaimer that is reachable, whilst allowing the reclaimer to perform both jobs in time. Suppose a third reclaim job is to be placed in between the existing two (only its loading period has to be in the reclaimer gap), as illustrated by rectangle in red. Whenever a job is assigned to a reclaimer, it restricts the other reclaimers movements (as they are mounted in the same rails). Also, because the reclaimer has to load the assigned stockpile, it would no longer be able to reach any other positions on the pad during this period. The area in space and time ``blocked'' by scheduling a reclaim job is illustrated in blue. Throughout this work we define the flexibility lost due to assigning a reclaim job as the area in space time that is reachable by a reclaimer but will no longer be reachable if the reclaim job is scheduled. In the figure above, that is the intersection between the reclaimer gap and the area that becomes unreachable by a reclaimer: $X+Y+Z$, where $Z$ is associated with the other reclaimer and equals $X$ or $Y$, depending on its position relative to the reclaimer under consideration. See \cite{Savelsbergh2014} for a more detailed explanation of flexibility loss.}
\label{fig:flexloss}
\end{figure}

When a reclaimer is assigned to a stockpile, we must make sure the reclaimer is able to reach the stockpile in time. After completing a reclaim job at a certain position of the pad, and assuming that reclaimers travel at a constant speed and that a reclaimer positions itself in the centre of the stockpile it is serving, the locations reachable by a reclaimer can be represented in time and space as a parallelogram, with opposing vertices connecting two reclaim jobs (one vertex at the end of a finishing reclaim job and the other at the beginning of the next reclaim job). This area, illustrated in pink in \autoref{fig:flexloss}, is referred to as \emph{reclaimer gap}. A collection of reclaimer gaps is a convenient algorithmic representation of the area reachable by a reclaimer at any time, and \autoref{algo:getReclGaps}() implements a function for this structure.

\begin{procedure}
\scriptsize
\caption{getReclaimerGaps($R, s$): Returns the set of all (adjusted) reclaimer gaps for reclaimer $R$ that can accommodate the reclaiming of stockpile $s$.}
\label{algo:getReclGaps}
\tcp{In our implementation we start with a single reclaimer gap connecting an artificial job ending $t=0$ at an arbitrary pad position, and ending at infinity (a sufficiently large number) at the same pad position, and split this gap appropriately as new reclaimer jobs are defined.}
\KwOut{A set of reclaimer gaps}
\end{procedure}

\begin{figure}[b]
\centering
\includegraphics[scale=0.55]{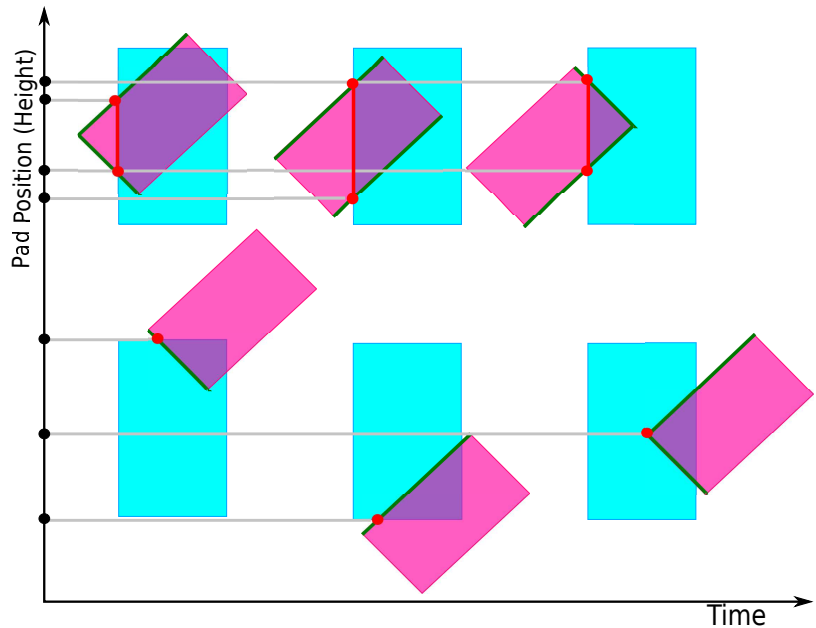}
\caption{Critical Heights. The above figure shows all possible ways a rectangle (the pad gap, in blue) and a parallelogram (the reclaimer gap, in magenta) can intersect. We are looking for earliest reclaim times, so we are interested in the leftmost points in the intersection. Unless the edge that represents the leftmost intersection between the pad and reclaimer gaps (represented in red) is a single point, there are infinitely many possible placements. In this case, only ``extreme'' placements (highlighted with red dots) are considered.}
\label{fig:criticalHeights}
\end{figure}

To be stacked and reclaimed a stockpile must be placed in a pad gap, and have its reclaim period inside a reclaimer gap. Therefore, feasible placements require that the pad and reclaimer gaps intersect, and reclaiming can start at any time in this intersection and can be placed in any position in it as well. Since we are interested in earliest reclaim times, we seek a point on the leftmost boundary of the intersection. Since there could be infinitely many points on this boundary, to obtain a finite set of positions to evaluate, we only consider extreme points on the boundary, which are referred to as \emph{Critical Heights}. \autoref{fig:criticalHeights} illustrates this concept, which is implemented by \autoref{algo:getCriticalHeights}(). 

\begin{procedure}
\scriptsize
\caption{getCriticalHeights($padGap, reclGap$): Returns a finite set of critical heights at which an earliest reclaim time can be found in the intersection of $padGap$ and $reclGap$.}
\label{algo:getCriticalHeights}
\tcp{In \autoref{fig:criticalHeights}}
Observe the relative positioning of the leftmost point of the reclaimer gap with respect to the leftmost edge of the pad gap\;
Identify the leftmost intersecting edges; \tcp{Represented in green}
Calculate the leftmost extreme points; \tcp{Represented as red dots}

\Return{The vertical positions of the red dots} \tcp{The black dots found on the vertical axis, duplicates excluded}
\Return{The vertical positions of reclaim jobs assigned to the other reclaimer that services the same pad} \tcp{Found by inspection}
\end{procedure}

Since two reclaimers serve each pad and they cannot pass each other, other than being able to reach a stockpile, we must also make sure a reclaimer does not clash with the other reclaimer. Once a reclaim job is assigned to a stockpile, the reclaimer remains stationary, positioned at the centre of stockpile, for the duration of the reclaim operation and therefore blocks any position that lies ``behind'' it. Assuming that, after a reclaim job, a reclaimer may move back to allow the other reclaimer to reach a position at that position (or further), the area in time and space that is blocked by a reclaimer can be represented as a trapezium, with its smaller base being the position of the stockpile being reclaimed (the blue area in \autoref{fig:flexloss}). \autoref{algo:getEarliestReclaimTime}() implements a function to find the earliest time at which reclaiming can start in this intersection and at the given critical height, accounting for possible reclaimer clashes.

\begin{procedure}
\scriptsize
\caption{getEarliestReclaimTime($stackEnd$, $padGap$, $reclGap$, $criticalHeight$):}
\label{algo:getEarliestReclaimTime}
Starting at $stackEnd$, find the earliest time in the intersection between $padGap$ and $reclGap$ minus the area blocked by the other reclaimer that services this pad that:
\begin{itemize}
	\item Allows enough reclaiming time to load the stockpile before the end of the pad or reclaimer gap, or the next forbidden time-space
	\item Has at least one ship loader available throughout the entire load period
	\item Has at least one berth available throughout the entire load period
\end{itemize}
\tcp{In our implementation this search is performed by sequential inspection}
\end{procedure}

\begin{procedure}
\scriptsize
\caption{comparePlacements(Stockpile Placement $sp1$, Stockpile Placement $sp2$): utility function that assigns a utility value to a stockpile placement.}
\label{algo:comparePlacements}
let $\underline{l_{s1}}$ and $\underline{l_{s2}}$ be the end of reclaiming times in $sp1$ and $sp2$ respectively\;
let $fl_1$ and $fl_2$ be the flexibility lost due to the positioning of the reclaimer in $sp1$ and $sp2$ respectively; \tcp{See \autoref{fig:flexloss}}
\uIf{$\lfloor \underline{l_{s1}} \rfloor \neq \lfloor \underline{l_{s2}} \rfloor$}{
	\lIf{$\lfloor \underline{l_{s1}} \rfloor < \lfloor \underline{l_{s2}} \rfloor$} {
		\Return{$sp1$ comes first}
	}
	\lElse{\Return{$sp2$ comes first}}
}
\uElseIf{$fl1 \neq fl2$}{
	\lIf{$fl1 < fl2$} {
		\Return{$sp1$ comes first}
	}
	\lElse{\Return{$sp2$ comes first}}
}
\Else{
	\lIf{$\underline{l_{s1}} < \underline{l_{s2}}$} {
		\Return{$sp1$ comes first}
	}
	\lElse{\Return{$sp2$ comes first}}
}
\end{procedure}

After enumerating every possible placement, \autoref{algo:slars}() selects the best one according to the comparator detailed by \autoref{algo:comparePlacements}. This would intuitively be the one that yields the earliest reclaim end time. However, we use an improved version proposed by \cite{Savelsbergh2014}, implemented by the \ref{algo:comparePlacements}, that considers the future placement flexibility lost due to possible reclaimer clashes as a ``look-ahead'' (illustrated by \autoref{fig:flexloss}), since \cite{Savelsbergh2014} found that it substantially improved the solutions.

For further information on pad and reclaimer gaps, critical heights and flexibility losses, we refer the reader to \cite{Savelsbergh2014}.

\subsection{Obtaining Improved Solutions Using a Genetic Algorithm}
\label{sec:ga}

Since the main engine we use to obtain good solutions - the algorithm detailed in the previous sections - is a greedy construction heuristic, it bases its decisions on the current state of the solution, e.g., used capacities and allocation of resources. Therefore, the order in which vessels are scheduled impacts the final average delay.

Aiming to increase the search space, to add diversity to our pool of explored
solutions, and ultimately to find better schedules, we propose a Genetic
Algorithm (GA) that exploits this observation, by exploring different vessel
scheduling orders. Since \ref{algo:slars}() is deterministic, an ordering maps
to exactly one solution. Under the assumption that good quality solutions
share common placements, the proposed GA attempts to preserve interesting
schedule traits (that is good subsequences), while proposing changes in different parts.

\subsubsection{Genetic algorithm}

A GA is a population-based search method that uses principles also found in the theory of evolution to find solutions for complex computational problems~\citep{Goldberg2014}. In general, the method starts with a diverse set of solutions - the population - which is then ``evolved'' through the use of particular operators, i.e., crossover, mutation, fitness calculation, selection and insertion, towards better solutions. The method does not guarantee global optimality but, if designed appropriately, it often leads to high quality solutions in shorter CPU times than required by exact methods. \autoref{algo:genetic} illustrates the concept.

\begin{algorithm}
\scriptsize
$population \leftarrow$ \ref{algo:initializePopulation}($numberOfSequences$)\;
\For(\texttt{//main loop}){$nGenerations \in \{1..maxGenerations\}$} { 
	\ref{algo:updatePopulationHierarchy}(); \tcp{keeps the population hierarchy correct}
		\tcp{Each iteration involves solving an independent SLARS sub problem}
		\ForEach(\texttt{//generation loop - pairs (parent, child) are evenly distributed amongst all free processors}){pair $(parent, child) \in population$} {
			$newSeq \leftarrow$ \ref{algo:crossover}($parent$, $child$)\;
			\ref{algo:mutate}($newSeq$)\;
			\ref{algo:calculateFitness}($newSeq$) \tcp{Solve a SLARS sub problem}
			\ref{algo:insert}($newSeq$)
		}
		$populationConverged \leftarrow$ \ref{algo:checkPopulationConvergence}($population$)\;
		\If {($populationConverged$)}{
			\ref{algo:populationRestart}($population$)\;
		}
}
reportBestSolution()\;
\caption{Genetic algorithm to define the order in which the vessels are scheduled.}
\label{algo:genetic}
\end{algorithm}

The first line of \autoref{algo:genetic} represents the creation of an initial
population of random sequences and assigning an objective function value
(fitness) to each of them by calling \ref{algo:slars}() for each sequence. In
the next line, the algorithm enters its main loop, which runs for a
pre-specified number of times. The random initial population is strongly
clustered around the turn-of-arrival ordering to allow the GA to converge in a
reasonable amount of time.

The main loop starts by calling a procedure to update the population structure. We use a single, small population organised in a ternary heap in which crossover only happens between adjacent nodes. This population structure helps to keep the number of individuals (and therefore, calls of \ref{algo:slars}()) low, and enforces convergence \citep{Buriol2004}. That procedure is followed by the generation loop, which comprises four methods called in sequence: crossover, mutation, fitness calculation and insertion. Each generation iteration generates 16 independent new SLARS subproblems (one for each physical core on our test machine). These are evenly distributed amongst all free available processors and solved independently before the tree is updated.

After a generation ends, a convergence check is performed to avoid spending too much time evolving a population composed of sequences that are too similar. If it has converged, the algorithm triggers an \emph{elitist} restart procedure: it recreates the population at random, but preserving the sequence that led to the best known solution. When the algorithm reaches its maximum number of generations, it stops and reports the best solution found so far.

\begin{procedure}
\scriptsize
\caption{initialisePopulation(numberOfSequences): a population has 16 solutions (one for each physical core on our test machine). Therefore the initialisation procedure creates 16 random sequences of integers involving numbers 1 through the number of vessels.}
\label{algo:initializePopulation}
\Repeat{$numberOfSolutions$ times} {
	$newSeq \leftarrow \ref{algo:createRandomSequence}(50\%)$\;
	\ref{algo:calculateFitness}($newSeq$) \tcp{Solve a SLARS sub problem}
	$population \leftarrow population \cup newSeq$\;
}
\end{procedure}

\begin{procedure}
\scriptsize
\caption{createRandomSequence(Probability $p$): reorders a sequence by swapping adjacent positions with a certain probability. This aims to keep variations small and localised around the turn of arrival.}
\label{algo:createRandomSequence}
Start with the ETA sequence (increasing order) $Sequence$\;
\lForEach{$idx \in Sequence$} { p\% chance to swap $idx$ with the next position }
\end{procedure}

\begin{procedure}
\scriptsize
\caption{calculateFitness(sequence): The fitness of a vessel sequence is the objective function value obtained by \ref{algo:slars}() using it as a parameter.}
\label{algo:calculateFitness}
$fitness \leftarrow \ref{algo:slars}(sequence)$ \tcp{Calculated from the vessel schedules, using \autoref{eq:delay}}
\end{procedure}

\begin{figure}[t!]
\centering
\includegraphics[scale=0.45]{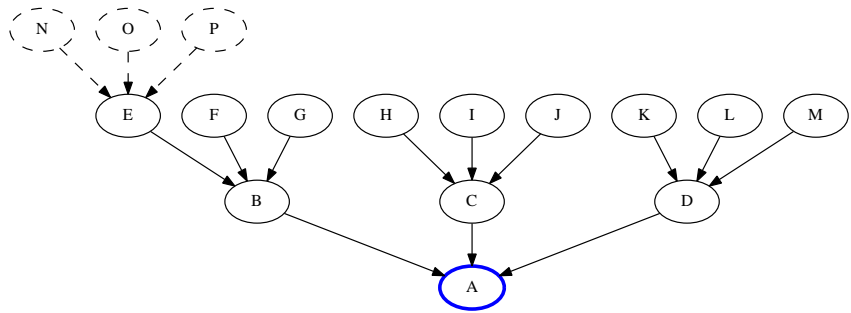}
\caption{The population hierarchy. The population is structured as a ternary tree, where the parent node always has a better objective function than its child nodes, making $A$ always the best solution in the population. Also, a crossover only happens between immediately connected nodes. In the above graph, each node represents a solution in the population, and the edges represent the relationship between them, i.e., assuming that the operator $<$ compares the fitness (average vessel delay) of two individuals, $A < v \in \{B,C,D\}$, $B < v \in \{E,F,G\}$ , $C < v \in \{H,I,J\}$ and $D < v \in \{K,L,M\}$. That is, a  crossover happens between $A$ and $B$, between $B$ and $E$, but \emph{not} between $A$ and $E$. We use a population with 16 individuals, to match the number of available processors in our test machines. A complete ternary tree would have 13 individuals. The extra three used to fill in the extra processors are depicted with dashed lines.}
\label{fig:ga_population}
\end{figure}

\begin{procedure}
\scriptsize
\caption{updatePopulationHierarchy(): Population hierarchy, structure and parent selection. The genetic algorithm uses a hierarchical population structured as a ternary heap. For our tests, given that each solution evaluation is time-consuming, it is critical to keep the number of individuals and fitness evaluations low, because each of them requires a call to \autoref{algo:slars}(). Therefore, we settled for a 16-individual population (one individual per physical core on our test machine) corresponding to an incomplete ternary tree with 4 levels. The population structure is hierarchical, with the parent nodes always having a better objective function (or fitness) than its three children. Therefore, the sequence that led to the best solution will always occupy the root node. In addition, crossover can only happen between adjacent nodes of the tree (i.e. between a parent node and one of its children nodes). \autoref{fig:ga_population} illustrates the tree structure. For a thorough discussion of the ternary tree structure used, as well as tests against unstructured and other structured population strategies, we refer the reader to~\cite{Buriol2004}.}
\label{algo:updatePopulationHierarchy}
\end{procedure}

\begin{procedure}
\scriptsize
\tcp{Let $Parent_A$ and $Parent_B$ be the parent sequences; $child$ be the new sequence}
$child \leftarrow \emptyset$; \tcp{Initialize $child$ as an empty sequence}
\tcp{Copy all vessels at the same position in both $Parent_A$ and $Parent_B$ to the $child$}
\For{$j = 1$ \textbf{\emph{to}} $numberOfVessels$} {
	\If {$Parent_A[j] == Parent_B[j]$} {
		$child[j] = Parent_A[j]$\;
	}
}

\tcp{Now complete the remaining empty positions of the $child$}
$index_A = index_B = index_{child}$ = 1; \tcp{indexes pointing to the first position of $Parent_A$ and $Parent_B$}
\While(\texttt{//find the first empty position of $child$}){$child[index_{child}] \neq \emptyset$} {
	$index_{child} = index_{child} + 1$\;
}

\While{$index_{child} \leq numberOfVessels$ } { \tcp{Only stops when $child$ is complete}
	$i$ = chooseRandomlyBetween($Parent_A$,$Parent_B$); \tcp{chooses one of the parents at random - 50\% chance to either}
	\While(\texttt{//finds the first vessel in $Parent_i$ that is not present in $child$}){$Parent_i[index_i] \in child$} {
		$index_i = index_i + 1$\;
	}
	\tcp{Here $index_i$ is pointing to a vessel in $Parent_i$ not yet added the $child$}
	$child[index_{child}] = Parent_i[index_i]$; \tcp{adds the vessel to the $child$}
	\While(\texttt{//finds the next empty position of $child$}){$child[index_{child}] \neq \emptyset$} {
		$index_{child} = index_{child} + 1$\;
	}
}
\Return{$child$}
\caption{crossover(Sequence 1, Sequence 2): Crossover between two sequences of vessels. Once a parent node and one of its children are selected, they will recombine to create a new sequence. The crossover will take the two sequences of integers (say A and B) and generate a third one (say C) by combining their information. Initially, the procedure copies all vessels that occupy the same position in A and B to C. This guarantees that a number of vessels will have their absolute positions preserved. Then, the procedure alternates randomly between A and B and copies the next unallocated vessel to C. This favours the relative order of the vessels, that is, if vessel $i$ is scheduled before vessel $j$ in A and B, it is likely (although not guaranteed) that the property is maintained in C. The crossover provides a good balance between maintaining important structures present in A and B and adding diversity to the offspring.}
\label{algo:crossover}
\end{procedure}

\begin{procedure}
\scriptsize
\caption{mutate(Sequence): implements the swap of two adjacent vessels in the sequence. The position is chosen at random.}
\label{algo:mutate}
\end{procedure}

\begin{procedure}
\scriptsize
\caption{insert(Sequence): a new sequence is always created from the crossover between a parent node and one of its children, where the parent node has a better fitness value than its children. Every time a new individual is created, if its fitness is better than that of the child node, it replaces it in the population tree, and is discarded otherwise. There is no acceptance of new individuals that are worse than their parents. That policy generates a strong evolutionary pressure which might lead to premature convergence, which is avoided by restarting the population when it is detected.}
\label{algo:insert}
\end{procedure}

\begin{procedure}
\scriptsize
\caption{checkPopulationConvergence(Population): Checking whether or not a population has converged is done under the assumption that as the population converges into very similar individuals, it will become more and more difficult to generate a child that is better than its parents. Following that, our procedure checks whether during an entire generation there was no new individual created that was better than its parents. When that happens, the population is assumed to have converged, triggering a restart procedure. This way, we minimise the time spent evolving individuals that are too similar, which most likely will not lead to any improvement of the incumbent solution.}
\label{algo:checkPopulationConvergence}
\end{procedure}

\begin{procedure}
\scriptsize
\caption{populationRestart(Population): When the population has converged, all sequences, except the one that led to the best known solution, are replaced by randomised ones, in the same way as the population initialisation process.}
\label{algo:populationRestart}
\end{procedure}


\section{Computational Results}
\label{sec:results}

To test the performance of our method, we used ten instances created with the stem generator developed by \cite{Boland2013}. An instance specifies a vessel arrival stream for a period of 100 days plus one year.  The first 100 days are to be used as a warm-up period to ensure that the system has reached its ``normal'' state before performance statistics are gathered. Due to the confidentiality agreement that is in place with our industry partners, we are not allowed to provide the instances or the parameters used to generate them. \autoref{tab:instances} summarises some of the instances characteristics.

\begin{table}
	\centering
	\begin{tabular}{lcccccccc}
	\multicolumn{3}{c}{} & \multicolumn{3}{c}{\textbf{ETAs (h)}} &\multicolumn{3}{c}{\textbf{Weight (t)}} \\
	\textbf{Inst} & \textbf{Vessels} & \textbf{Stockpiles} & \textbf{Min} & \textbf{Max}
 	& \textbf{Diff} & \textbf{Min} & \textbf{Max} & \textbf{Avg} \\
	\multicolumn{9}{c}{\textit{\textbf{KCT}}} \\
	\hline
	1 & 104 & 168 & 537.65 & 1373.2 & 8.11 & 10000 & 153310 & 67342.4 \\
	2 & 102 & 163 & 1376 & 2201.95 & 8.18 & 10000 & 160000 & 68252.81 \\
	3 & 101 & 162 & 2212.67 & 3054.86 & 8.42 & 10000 & 155021.97 & 71526.44 \\
	4 & 103 & 160 & 3085.83 & 3940.77 & 8.38 & 10000 & 160000 & 71996.09 \\
	5 & 109 & 160 & 3973.64 & 4791.08 & 7.57 & 10000 & 148242.01 & 71391.7 \\
	6 & 105 & 169 & 4795.06 & 5670.9 & 8.42 & 10000 & 153310 & 70898.35 \\
	7 & 109 & 162 & 5671.8 & 6509.51 & 7.76 & 10000 & 160000 & 72082.48 \\
	8 & 106 & 162 & 6525.48 & 7416.83 & 8.49 & 10000 & 153310 & 72486.98 \\
	9 & 101 & 164 & 7424.57 & 8256.45 & 8.32 & 10000 & 148596.49 & 69002.86 \\
	10 & 102 & 153 & 8257.64 & 9093.82 & 8.28 & 10000 & 150089 & 72211.86 \\
	\multicolumn{9}{c}{\textit{\textbf{CCT}}} \\
	\hline
	1 & 30 & 40 & 581.11 & 1286.81 & 24.33 & 10000 & 129288 & 61820.81 \\
	2 & 32 & 42 & 1378.85 & 2182.53 & 25.93 & 10000 & 136343 & 62587.97 \\
	3 & 28 & 34 & 2237.45 & 3075.64 & 31.04 & 10000 & 153310 & 68000.67 \\
	4 & 34 & 47 & 3120.18 & 3947.67 & 25.08 & 10000 & 99999 & 50409.16 \\ 
	5 & 27 & 39 & 3976.98 & 4725.46 & 28.79 & 10000 & 114572.01 & 56368.35 \\
	6 & 29 & 43 & 4841.54 & 5666.62 & 29.47 & 10000 & 111617 & 55923.77 \\
	7 & 32 & 45 & 5709.9 & 6515.56 & 25.99 & 10000 & 118563 & 57638.02 \\
	8 & 27 & 36 & 6525.1 & 7423.06 & 34.54 & 10000 & 137055 & 65048.63 \\
	9 & 34 & 50 & 7436.91 & 8244.56 & 24.47 & 16380 & 97950 & 47943.16 \\
	10 & 34 & 46 & 8256.89 & 9087.6 & 25.17 & 10000 & 93866 & 52910.53 \\
	\multicolumn{9}{c}{\textit{\textbf{NCT}}} \\
	\hline
	1 & 66 & 67 & 554.44 & 1356.58 & 4.2 & 16940 & 155579 & 93503.45 \\
	2 & 66 & 69 & 1379.09 & 2204.14 & 4.16 & 16940 & 144474 & 92295.32 \\
	3 & 71 & 74 & 2210.42 & 3050.83 & 4.35 & 16940 & 160000 & 88516.73 \\
	4 & 63 & 64 & 3089.35 & 3949.82 & 4.34 & 26940 & 160000 & 102664.05 \\
	5 & 64 & 66 & 3975.3 & 4789.99 & 4.11 & 26940 & 160000 & 95594.41 \\
	6 & 66 & 69 & 4792.38 & 5653.18 & 4.41 & 26940 & 160000 & 95832.85 \\
	7 & 59 & 61 & 5679.5 & 6507.72 & 4.24 & 26940 & 155579 & 96397.07 \\
	8 & 67 & 69 & 6519.67 & 7411.34 & 4.54 & 16940 & 160000 & 97557.11 \\
	9 & 65 & 69 & 7453.27 & 8254.33 & 4.18 & 26940 & 160000 & 97902.91 \\
	10 & 64 & 69 & 8257.4 & 9092.82 & 4.21 & 16940 & 155277 & 91992.16 \\
	\multicolumn{9}{c}{\textit{\textbf{System}}} \\
	\hline
	1 & 200 & 275 & 537.65 & 1373.2 & 4.2 & 10000 & 155579 & 72913.04 \\
	2 & 200 & 274 & 1376 & 2204.14 & 4.16 & 10000 & 160000 & 73438.97 \\
	3 & 200 & 270 & 2210.42 & 3075.64 & 4.35 & 10000 & 160000 & 75739.05 \\
	4 & 200 & 271 & 3085.83 & 3949.82 & 4.34 & 10000 & 160000 & 75494.85 \\
	5 & 200 & 265 & 3973.64 & 4791.08 & 4.11 & 10000 & 160000 & 75208.56 \\
	6 & 200 & 281 & 4792.38 & 5670.9 & 4.41 & 10000 & 160000 & 74729.57 \\
	7 & 200 & 268 & 5671.8 & 6515.56 & 4.24 & 10000 & 160000 & 75191.39 \\
	8 & 200 & 267 & 6519.67 & 7423.06 & 4.54 & 10000 & 160000 & 77962.85 \\
	9 & 200 & 283 & 7424.57 & 8256.45 & 4.18 & 10000 & 160000 & 72328.37 \\
	10 & 200 & 268 & 8256.89 & 9093.82 & 4.21 & 10000 & 155277 & 73991.64 \\
	\hline
	\end{tabular}
	\caption{Instance Summary. For each terminal and instance: the number of vessels, number of cargoes, the first and last ETAs, and the average difference between two consecutive ETAs; and the smallest, biggest and average cargo tonnages.}
	\label{tab:instances}
\end{table}

We provide and discuss results obtained with the GA detailed in \autoref{sec:ga} and with the Constraint Programming (CP) approach of \cite{Belov2014}. Their method is also designed to solve the problem described in \autoref{sec:problem}, and differs from our method in that it schedules the activities in NCT and CCT in more detail, i.e., it also decides stockpile locations and schedules stacking and reclaiming. Because of the smaller volume of coal handled by these terminals and because the bottlenecks of the logistics system are more likely to be related to transport capacity \citep{Belov2014}, the average vessel delays should be comparable.

Additionally, we consider results obtained with a simple randomized
Multi-Start (MS) heuristic to highlight the effectiveness the proposed GA
scheme. The MS algorithm simply runs \ref{algo:slars}() using random sequences
obtained in a similar manner as done in \autoref{algo:initializePopulation}(),
dispatching one such run to a CPU as it becomes available, and reports the
best solution found. A reduced chance of swapping positions, 30\% rather than
50\%, is used to ensure that the obtained sequences do not deviate too much
from the increasing ETA order, which is known to be good. Given the vast
search space, focussing the search in the vicinity of the chronological
ordering is essential for finding good solutions in a reasonable amount of
time. See \cite{singh2012mixed} where it was shown for a
similarly large problem arising from the same supply chain, that a GA without
such a targeted search is not competitive.

The tests with the GA and MS algorithms were ran on dual octa core 3.33GHz Intel Xeon E5-2667 v2 processors with 256GB RAM. For both these algorithms memory consumption was negligible (only a few MBs). The runs with CP were performed on an octa core 3.40GHz Intel Core i7-2600 with 8GB RAM and kindly provided by the original authors of \cite{Belov2014}. The authors reported a memory usage of around 150MB for CP. It is also noteworthy that CP runs on a single thread.

Since both GA and MS have a random component, the reported results are the average of ten runs. CP, on the other hand, is deterministic and a single run is reported.

In the following experiments, each run of the GA includes 100 generations. Since every generation calls \ref{algo:slars}() 16 times, in the following experiments 1600 iterations of MS are performed to match the number of \ref{algo:slars}() calls.

\begin{table}
	\centering
	\caption{The Average Vessel Delays of the system in hours. The numbers reported for each instance are the average of 10 runs with different seeds in the case of MS and GA, and a single run in the case of CP. MS is a simple Multi-Start procedure that calls \ref{algo:slars}() with random vessel orders. GA is the procedure detailed in \autoref{algo:genetic}. CP is the method proposed by \cite{Belov2014}. The results in italics show the best known solutions.}
	\begin{tabular}{lccccc}
		\hline
		& \multicolumn{2}{l}{\textbf{MS}} & \multicolumn{2}{l}{\textbf{GA}} & \textbf{CP} \\
		\textbf{Instance} & \textbf{Avg (h)} & \textbf{Std Dev} & \textbf{Avg (h)} & \textbf{Std Dev} & \textbf{Delay (h)} \\
		\hline
		\textbf{1} & 8.27 & 0.12 & \textit{7.37} & 0.25 & 7.61 \\
		\textbf{2} & 7.32 & 0.22 & \textit{6} & 0.13 & 8.16 \\
		\textbf{3} & 6.49 & 0.1 & \textit{5.64} & 0.13 & 7.12 \\
		\textbf{4} & 9.35 & 0.1 & \textit{8.42} & 0.24 & 9.47 \\
		\textbf{5} & 8.79 & 0.12 & \textit{7.85} & 0.14 & 11.67 \\
		\textbf{6} & 5.37 & 0.12 & \textit{4.89} & 0.1 & 5.17 \\
		\textbf{7} & 7.94 & 0.19 & \textit{7.18} & 0.19 & 8.59 \\
		\textbf{8} & 6.44 & 0.05 & \textit{5.87} & 0.22 & 6.7 \\
		\textbf{9} & 9.1 & 0.19 & \textit{7.61} & 0.13 & 9.62 \\
		\textbf{10} & 6.34 & 0.06 & \textit{5.73} & 0.16 & 6.24 \\
		\hline
	\end{tabular}
	\label{tab:avgdelays}
\end{table}

\begin{table}
	\centering
	\caption{Running Times. The numbers reported for each instance are the average of 10 runs with different seeds in the case of MS and GA, and a single run in the case of CP. MS is a simple Multi-Start procedure that calls \ref{algo:slars}() with random vessel orders. GA is the procedure detailed in \autoref{algo:genetic}. CP is the method proposed by \cite{Belov2014}.}
	\begin{tabular}{lccccc}
		\hline
		& \multicolumn{2}{l}{\textbf{MS}} & \multicolumn{2}{l}{\textbf{GA}} & \textbf{CP} \\
		\textbf{Instance} & \textbf{Avg (s)} & \textbf{Std Dev} & \textbf{Avg (s)} & \textbf{Std Dev} & \textbf{Time (s)} \\
		\hline
		\textbf{1} & 85.38 & 0.31 & 226.22 & 10.65 & 6073.8 \\
		\textbf{2} & 77.67 & 0.27 & 195.74 & 14.13 & 7308.9 \\
		\textbf{3} & 81.9 & 1.37 & 208.3 & 8.76 & 6743.6 \\
		\textbf{4} & 81.9 & 1.83 & 195.37 & 8.34 & 9068.5 \\
		\textbf{5} & 81.94 & 0.94 & 213.77 & 18.14 & 9552.2 \\
		\textbf{6} & 81.94 & 1.51 & 236.58 & 16.76 & 6258.9 \\
		\textbf{7} & 77.04 & 1.51 & 197.75 & 11.06 & 9063.8 \\
		\textbf{8} & 78.65 & 1.39 & 202.82 & 10.21 & 7487.5 \\
		\textbf{9} & 80.94 & 1.24 & 192.32 & 8.98 & 7890.2 \\
		\textbf{10} & 72.66 & 1.4 & 201.96 & 11.58 & 6720.2 \\
		\hline
	\end{tabular}
	\label{tab:runtimes}
\end{table}

In \autoref{tab:avgdelays} we can see that GA also always provided better solutions than CP. This is possibly a result of the fact that GA considers a simplified model for CCT and NCT. Even though less restrictive, our model should be appropriate because most of the operational decisions at these terminals are made by the (independent) controllers of the companies that own specific terminal stockyard space, and little is known about their control strategies. The results in \cite{Belov2014} also show that operations at CCT and NCT are not very restrictive for the system and that the biggest bottleneck for optimising the system are operations related to the channel. On the other hand, our model does not discretise time, which could lead to a better use (in time and space) of the pads at KCT, and the proposed GA always considers all vessels as a group, while CP optimises over a rolling visibility window of 15 vessels at a time, which may allow GA to explore options not available to CP.

The results reported in \autoref{tab:runtimes} clearly show that both MS and GA vastly outperform CP in running time, finishing 100 generations of GA and 1600 re-starts of MS within a few minutes while CP often required a few hours of computation time. The significant improvement in terms of running times is due to a combination of the efficient implementation of the SLARS solver, and a fast convergence parallel GA framework. The fact that the average delays obtained by the GA are significantly lower than those obtained with a random multi-start procedure, support the fact that the GA was actually effective at guiding the search towards high quality solutions. Because the proposed model does not discretise time, relatively few decision variables are considered. The enumeration tree under consideration is also aggressively pruned by the hierarchical structure of the population, and crossover operator, which considers a consensus between arguably good solutions. Finally, the use of parallelism also helped to speed up the generation loops.

Looking further in \autoref{tab:runtimes}, even though GA required twice the time to solve the same number of problems as MS, the results are always better. The bigger running times are due to the fact that GA has a synchronisation step at the end of each generation, to ensure that the solution hierarchy is consistent for that generation, never containing individuals who belong to previous or following generations. The significant decrease in average delay, however, suggests that the GA was successful in exploiting common interesting solution characteristics to guide the search towards better solutions, and not wasting iterations solving sub-problems that were already solved before.

\begin{figure}
	\begin{subfigure}[b]{0.5\textwidth}
		\includegraphics[scale=0.58]{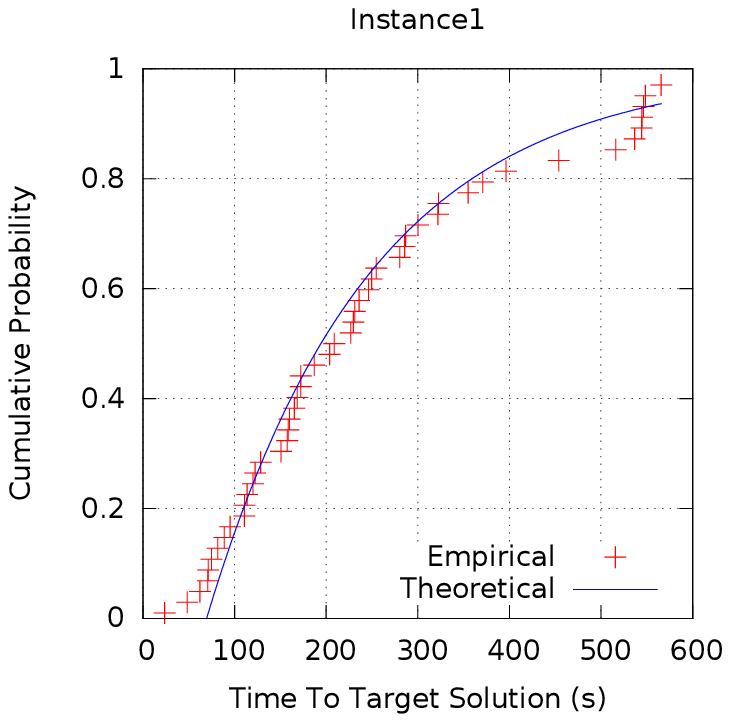}
	\end{subfigure}
	\begin{subfigure}[b]{0.5\textwidth}
		\includegraphics[scale=0.58]{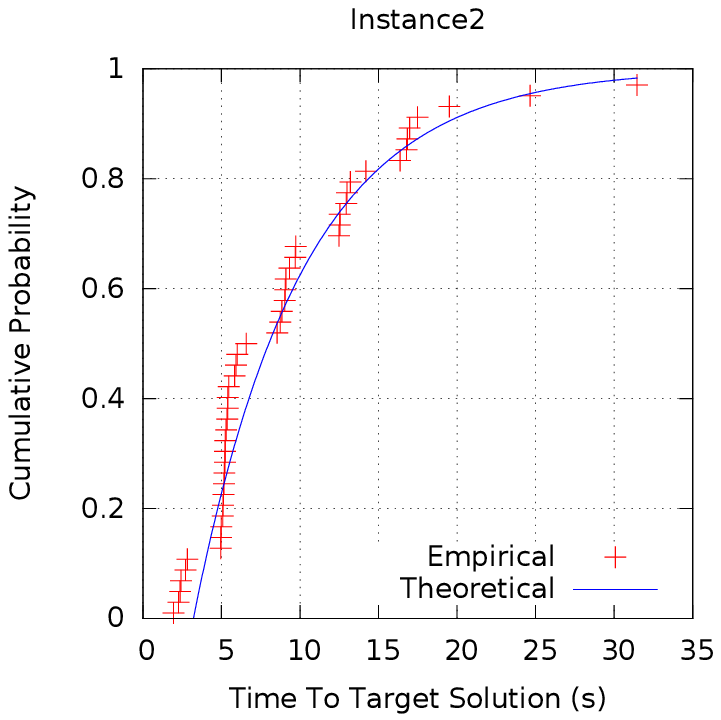}
	\end{subfigure}
	\begin{subfigure}[b]{0.5\textwidth}
		\includegraphics[scale=0.58]{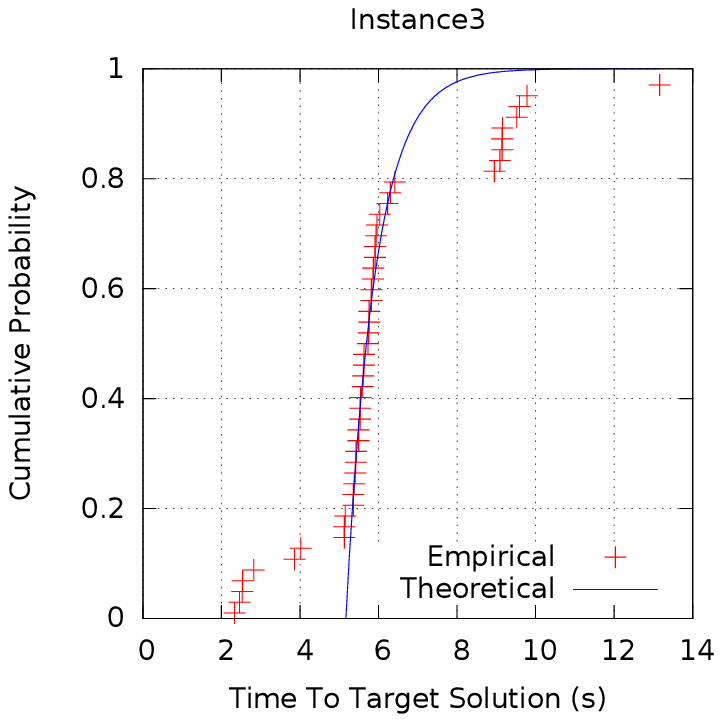}
	\end{subfigure}
	\begin{subfigure}[b]{0.5\textwidth}
		\includegraphics[scale=0.58]{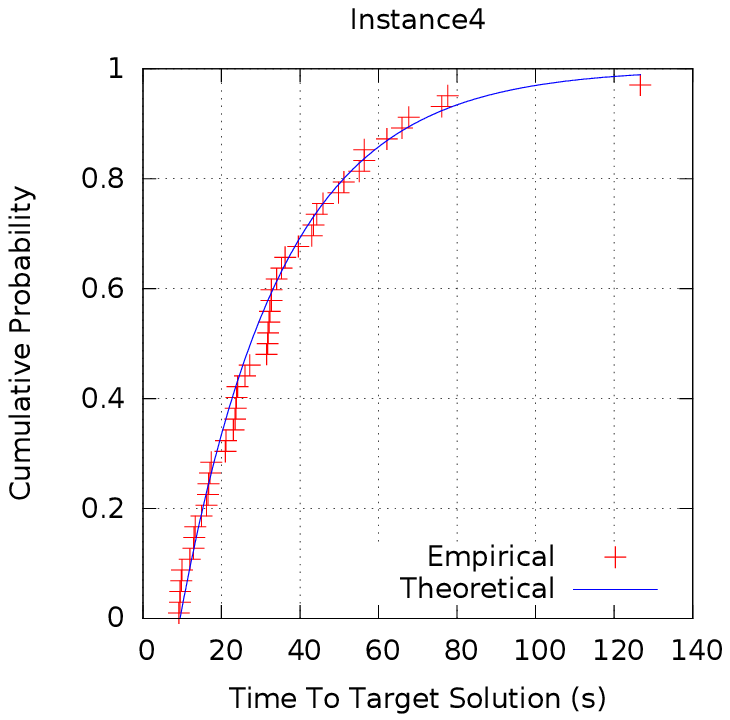}
	\end{subfigure}
	\begin{subfigure}[b]{0.5\textwidth}
		\includegraphics[scale=0.58]{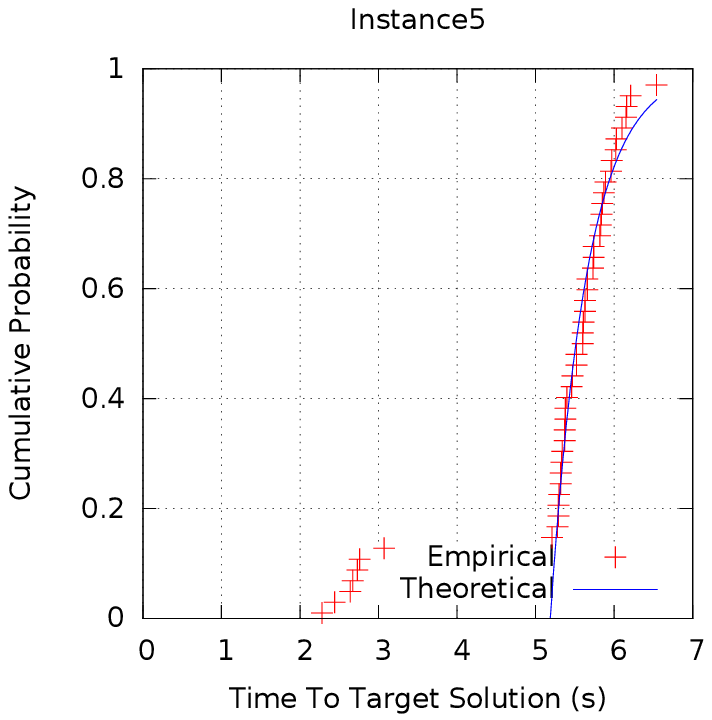}
	\end{subfigure}
	\begin{subfigure}[b]{0.5\textwidth}
		\includegraphics[scale=0.58]{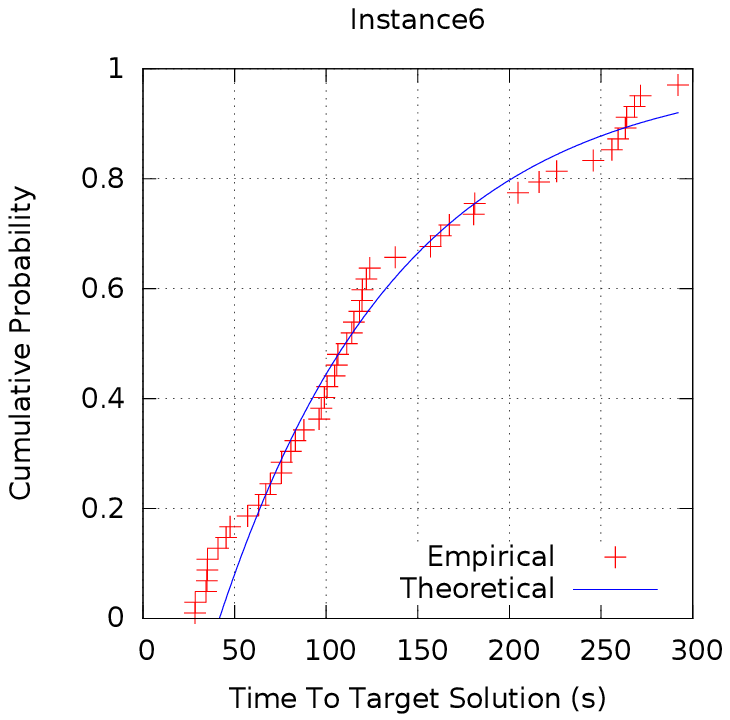}
	\end{subfigure}
	\begin{subfigure}[b]{0.5\textwidth}
		\includegraphics[scale=0.58]{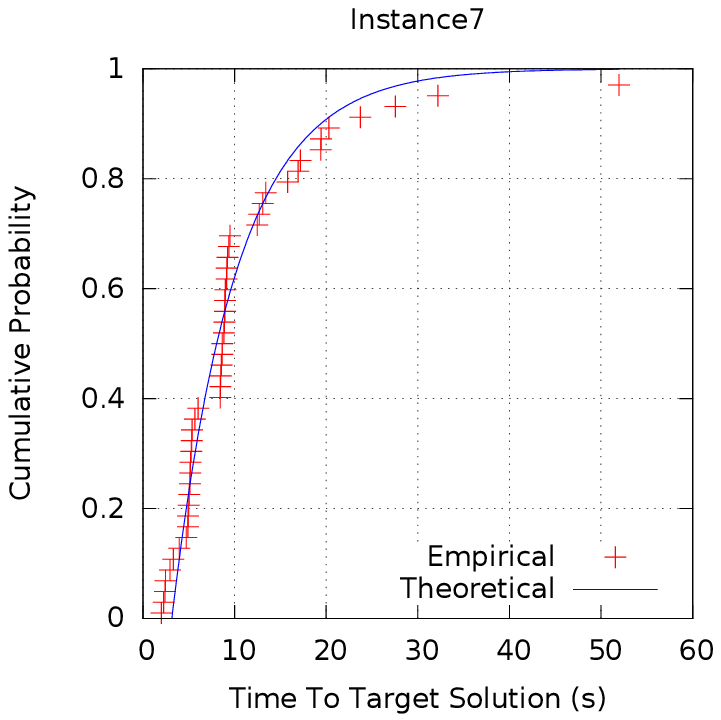}
	\end{subfigure}
	\begin{subfigure}[b]{0.5\textwidth}
		\includegraphics[scale=0.58]{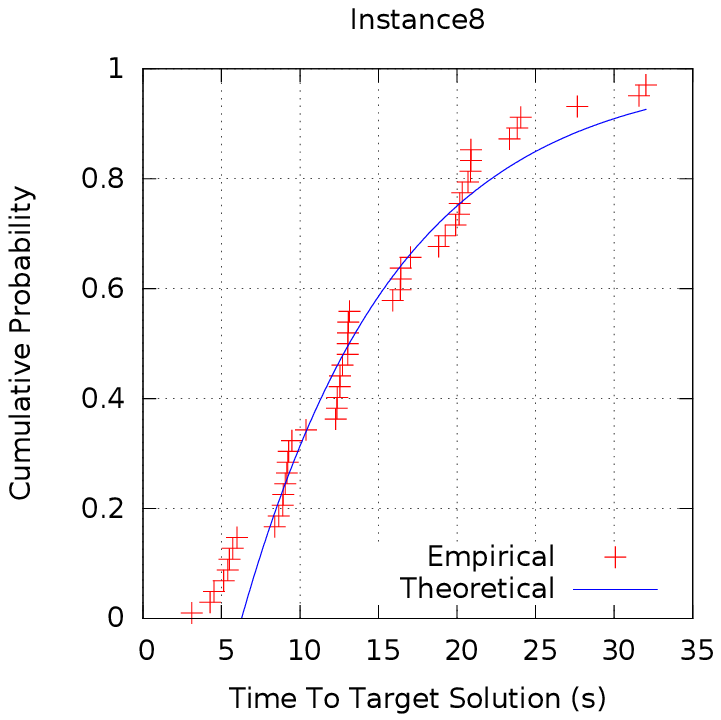}
	\end{subfigure}
	\begin{subfigure}[b]{0.5\textwidth}
		\includegraphics[scale=0.58]{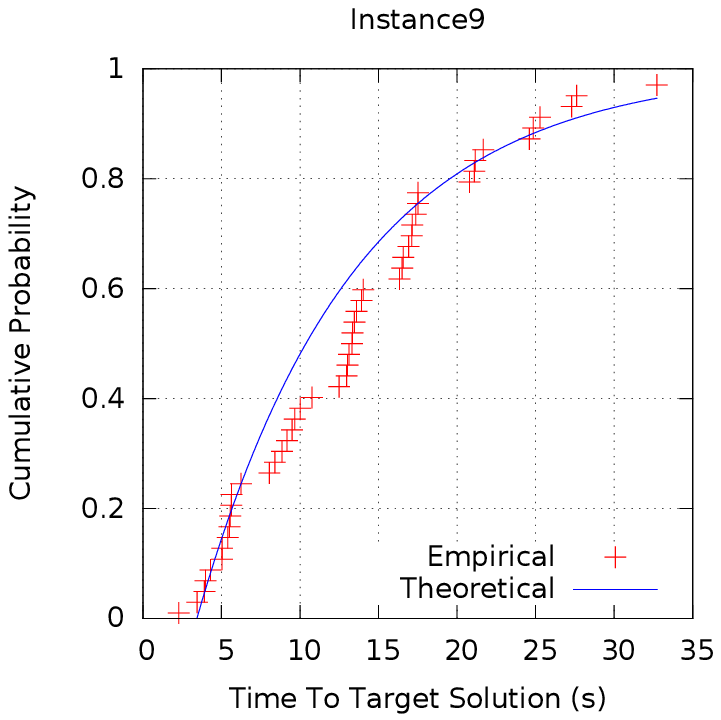}
	\end{subfigure}
	\begin{subfigure}[b]{0.5\textwidth}
		\includegraphics[scale=0.58]{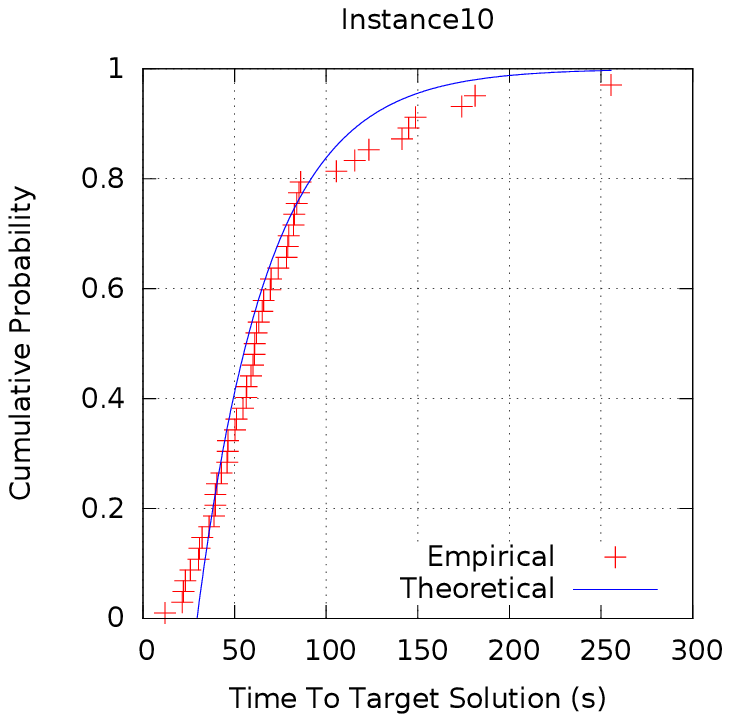}
	\end{subfigure}
\caption{Time to Target plots. The above figure shows the cumulative probabilities that the GA will reach or outperform the average vessel delays obtained by CP (as shown on \autoref{tab:avgdelays}), for each of the instances in our test bed. These follow the methodology proposed by \cite{Resende-2007}, and depict 50 runs of the GA for each instance.}
\label{fig:ttt}
\end{figure}

\autoref{fig:ttt} illustrates the probable times that GA will need to match CP's results using Time-To-Target plots (TTTplots, as proposed by \cite{Resende-2007}) generated with results from 50 runs for each of the instances, with the target set to that obtained by CP (as seen on \autoref{tab:avgdelays}). \autoref{fig:ttt} shows that GA matches or provides a better solution in every instance of our test bed, using only a fraction of the time required by CP.

\section{Conclusions}
\label{sec:conclusions}

In this work, we propose high performance, cost effective (in the sense that they do not rely on external solvers), algorithm to simultaneously schedule train and vessel arrivals and stockpile build and load periods in a system with three coal export terminals, whilst also positioning the stockpiles on specific pads and scheduling the reclaimers for the largest terminal. Given the fast run times and the modeling detail, it can be used by HVCCC to support both strategic and tactical decision making, allowing the analysis of various ``what if'' scenarios. Such scenarios would include different resource capacities and, more importantly, different stems (vessel arrival streams). The study of the obtained solution helps identify current (or future) bottlenecks in the coal chain and its operations, insight in how to improve throughput with minimal investment with a quantitative estimate of added value.

Our carefully implemented and tuned method outperforms the CP approach of \citet{Belov2014}, the best-performing method up to now, improving the best-known solutions for every instance in our test bed in less than one tenth of the time used by the CP approach. 

Possible further work includes the development of more efficient asynchronous parallelisation strategies, to reduce waiting periods, and more intensive searches on pad placement, e.g., exploring different suboptimal placements during the greedy procedure, which may help to make convergence to better solutions faster.

\bibliographystyle{spbasic}
\bibliography{refs}

\end{document}